\RequirePackage{fix-cm}
\documentclass[smallextended]{svjour3}       % onecolumn (second format)
\smartqed  % flush right qed marks, e.g. at end of proof
\usepackage[utf8]{inputenc}
\usepackage{graphicx}
\usepackage{alltt}
\usepackage{amsfonts}
\usepackage{url}
\newcommand{\denote}[1]{[\![#1]\!]}

\begin{document}

\title{Distant decimals of $\pi$ \thanks{This work was partially funded by the ANR project \emph{FastRelax}(ANR-14-CE25-0018-01) of the French National Agency for Research}
}

\subtitle{Formal proofs of some  algorithms computing them and guarantees of exact computation}


\author{Yves Bertot         \and
        Laurence Rideau \and
        Laurent Th\'ery %etc.
}


\institute{Yves Bertot, Laurence Rideau, Laurent Th\'ery \at
              Inria Sophia Antipolis \\
              Université Côte d'Azur\\
              \email{firstname.name@inria.fr}           %  \\
}

\date{Received: date / Accepted: date}

\maketitle

\begin{abstract}
We describe how to compute very far decimals of \(\pi\) and how to provide
formal guarantees that the decimals we compute are correct.  In particular,
we report on an experiment where 1 million decimals of \(\pi\)
and the billionth hexadecimal (without the preceding ones) have been computed
in a formally verified way.  Three methods have been studied, the first one
relying on a spigot formula to obtain at a reasonable cost only one
distant digit (more precisely a hexadecimal digit, because the
numeration basis is 16) and the other two relying on 
\emph{arithmetic-geometric means}.
All proofs and computations can be made inside the Coq system.  We detail
the new formalized material that was necessary for this achievement and the
techniques employed to guarantee the accuracy of the computed digits, in spite
of the necessity to work with fixed precision numerical computation.
\keywords{Formal proofs in real analysis \and Coq proof assistant \and
  Arithmetic Geometric Means \and Bailey \& Borwein \& Plouffe formula
  \and BBP \and PI}
\end{abstract}

\section{Introduction}
\label{intro}
The number \(\pi\) has been exciting the curiosity of mathematicians for
centuries.  Ingenious formulas to compute this number manually were
devised since antiquity with Archimede's exhaustion method and a notable
step forward achieved in the eighteenth century, when John Machin devised
the famous formula he used to compute one hundred decimals of \(\pi\).

Today, thanks to electronic computers, the representation of \(\pi\) in
fractional notation is known up to tens of trillions of decimal
digits.  Establishing such records raises some questions.  How do we know
that the digits computed by the record-setting algorithms are correct?
The accepted approach is to perform two computations using two different
algorithms.  In particular, with the help of a spigot formula, it is
possible to perform a statistical verification, simply checking that
a few randomly spread digits are computed correctly.

In this article, we study the best known spigot formula, an algorithm
able to compute a faraway digit at a cost that is much lower than
computing all the digits up to that position.  We also study two
algorithms based on arithmetic geometric means, which are based on
iterations that double the number of digits known at each step.  For these
algorithms, we perform all
the proofs in real analysis that
show that they do converge towards \(\pi\), giving
the rate of convergence, and we then show that all the
computations in a framework of fixed precision computations, where
computations are only approximated by rational numbers with a fixed
denominator, are indeed correct, with a formally proved bound on the
difference between the result and \(\pi\).  Last we show how we implement
the computations in the framework of our theorem prover.

The first algorithm, due to Bailey, Borwein, and Plouffe relies on a formula of the following shape, known as the BBP formula \cite{BBP97}.

\[\pi = \sum_{i=0}^{\infty} \frac{1}{16 ^ i}  \left(\frac{4}{8 i + 1} -
\frac{2}{8 i + 4}  - \frac{1}{8 i + 5} - \frac{1}{8 i + 6}\right). \]

Because each term of the sum is multiplied by \(\frac{1}{16^i}\) it
appears that
approximately \(n\) terms of the infinite sum are needed to compute the
value of the
nth hexadecimal digit.  Moreover, if we are only interested in the value of the
nth digit, the sum of terms can be partitioned in two parts, where
the first contains the terms such that \(i \leq n\) and the second contains
terms that will only contribute when carries need to be propagated.

We shall describe how this algorithm is proved correct and what
techniques are used to make this algorithm run inside the Coq theorem prover.

The second and third algorithms rely on a process known as the
\emph{arithmetic-geometric mean}.  This process considers two inputs
\(a\) and \(b\) and successively computes two sequences \(a_n\) and 
\(b_n\) such that \(a_0 = a\), \(b_0 = b\), and
\[a_{n+1} = \frac{a_n + b_n}{2} \qquad b_{n+1} = \sqrt{a_n b_n}\]

In the particular case where \(a = 1\) and \(b = x\), the values
\(a_n\) and \(b_n\) are functions of \(x\) that are easily shown to be
continuous and differentiable and it is useful to consider the two functions
\[y_n (x) = \frac{a_n(x)}{b_n(x)} \qquad z_n = \frac{b'_n(x)}{a'_n(x)}\]
A first computation of \(\pi\) is expressed by the following equality:
\[ \pi = (2 + \sqrt{2}) \prod_{n = 1} ^{\infty} \frac{1 + y_n(\frac{1}{\sqrt{2}})}
{1 + z_n(\frac{1}{\sqrt{2}})}.\]
Truncations of this infinite product are shown to approximate \(\pi\)
with a number of decimals that doubles every time a factor is added.
This is the basis for the second algorithm.

The third algorithm also uses the arithmetic geometric mean for 
\(1\) and \(\frac{1}{\sqrt{2}}\), but performs a sum and a single
  division:
\[\pi = \lim_{n\rightarrow \infty} \frac{4 (a_n(1, \frac{1}{\sqrt{2}}))^2}
{1 - \sum_{i=1}^{n-1} 2 ^ {i - 1} (a_{i-1}(1,\frac{1}{\sqrt{2}}) -
  b_{i - 1}(1,\frac{1}{\sqrt{2}})) ^ 2}\]
It is sensible to use index \(n\) in the numerator and \(n-1\) in the sum
of the denominator, because this gives approximations with comparable
precisions of their respective limits.  This is the basis for the third
algorithm.  This third algorithm was introduced in 1976 independently
by Brent and Salamin \cite{Brent76,Salamin76}.  It is the one
implemented in the {\tt mpfr}
library for high-precision computation \cite{mpfr} to compute \(\pi\).

In this paper, we will recapitulate the mathematical proofs of these
algorithms (sections~\ref{sec:plouffe} and~\ref{sec:agm}), and show what
parts of existing libraries of real analysis
we were able to reuse and what parts we needed to extend.

For each of the algorithms, we study first the mathematical
foundations, then we concentrate on implementations where all
computations are done with a single-precision fixed-point arithmetic,
which amounts to
forcing all intermediate results to be rational numbers with a common
denominator.  This framework imposes that we perform more proofs
concerning bounds on accumulated rounding errors.
\paragraph{Context of this work.}  All the work described in this
paper was done using the Coq proof assistant \cite{coq}.  This system
provides a library describing the basic definition of real analysis,
known as {\em the standard Coq library for reals}, where the existence
of the type of real numbers as an ordered, archimedian, and complete
field with decidable comparison is assumed.  This choice of foundation
makes that mathematics based on this library is inherently classical,
and real numbers are abstract values which cannot be exploited in
the programming language that comes in Coq's type theory.

The standard Coq library for reals provides notions like convergent
sequences, series, power series, integrals, and derivatives.  In
particular, the sine and cosine functions are defined as power series,
\(\pi\) is defined as twice the first positive root of the cosine
function, and the library provides
a first approximation of \(\frac{\pi}{2}\) as being between
\(\frac{7}{8}\) and \(\frac{7}{4}\).  It also provides a formal
description of Machin formulas, relating computation of \(\pi\) to
a variety of computations of arctangent at rational arguments, so that
it is already possible to compute relatively close approximations of
\(\pi\), as illustrated in \cite{BertotAllais14}.

The standard Coq library implements principles that were designed
at the end of the 1990s, where values whose existence is questionable
should always be guarded by a proof of existence.  These principles
turned out to be impractical for ambitious formalized mathematics
in real analysis, and a new library called Coquelicot
\cite{BLM15} was designed
to extend the standard Coq library and achieve a more friendly 
and regular interface for most of the concepts, especially limits,
derivatives, and integrals.  The developments described in this
paper rely on Coquelicot.

Many of the intermediate level steps of these proofs are performed
automatically.  The important parts of our working context in this
respect are the {\tt Psatz} library, especially the {\tt psatzl}
tactic \cite{DBLP:conf/types/Besson06}, which solves reliably all
questions that can be described
as linear arithmetic problems in real numbers and {\tt lia}
\cite{DBLP:conf/types/Besson06}, which
solves similar problems in integers and natural numbers.  Another
tool that was used more and more intensively during the development
of our formal proofs is the {\tt interval} tactic \cite{interval},
which uses interval arithmetic to prove bounds on mathematical
formulas of intermediate complexity.  Incidentally, the {\tt interval}
tactic also provides a simple way to prove that \(\pi\) belongs
to an interval with rational coefficients.

Intensive computations are performed using a library for computing
with very large integers, called {\tt BigZ}~\cite{GregoireTheryIJCAR2006}.
It is quite notable that this library contains an implementation of
an optimized algorithm to compute square roots of large
integers~\cite{RacineCarreeGMP}.

\section{The BBP formula}
\label{sec:plouffe}
In this section we first recapitulate the main mathematical formula
that makes it possible to compute a single hexadecimal at a low cost
\cite{BBP97}.

Then, we describe an implementation of an algorithm that performs the
relevant computation and can be run directly inside the Coq theorem
prover.
\subsection{Proof of the BBP formula}
\label{sec:plouffe_proof}

%%  \[\sum_{i=0}^{\infty} \frac{1}{16 ^ i} \left(4\frac{1}{8 i + 1} -2
%%  \frac{1}{8 i + 4}  - \frac{1}{8 i + 5} - \frac{1}{8 i + 6}\right) = \int_{0}^{\frac{1}{\sqrt{2}}} \frac{4\sqrt{2} - 8 x^3 -4\sqrt{2}x^4 -8x^5} { 1 - x^8} dx \]

%% \[= \int_{0}^1 \frac {16(y -1)}{y^4 - 2 y^3 +4y -4}dy\]
%% \[= \int_{0}^1 4\frac{2-y}{y^2 -2y +2}dy + \int_{0}^1 4\frac{y}{y^2 -2}dy \]
%% \[ =  \int_{0}^1 \frac {4 -4y}{y^2 -2y +2}dy + \int_{0}^1\frac{4}{1 + (y -1)^2}dy + \int_{0}^1 4\frac{y}{y^2 -2}dy\]
%% \[ = \left[-2ln(y^2 -2y +2)+4arctan(y-1)+2ln(2-y^2)\right]_0^1\]
%% \[=\pi\]

\subsubsection{The mathematical Proof}
We give here a detailed  proof of the formula established by David Bayley, Peter Borwein and Simon Plouffe.  The level of detail is chosen to mirror the difficulties encountered in the formalization work.
\begin{equation}\label{eqn:plouffe}
\pi = \sum_{i=0}^{\infty} \frac{1}{16 ^ i} (\frac{4}{8 i + 1} -\frac{2}{8 i + 4}  - \frac{1}{8 i + 5} - \frac{1}{8 i + 6})
\end{equation}
We first study the properties of the sum $S_k$ for a given $k$ such that $1 < k $:
\begin{equation}\label{eqn:SK}
S_k = \sum_{i=0}^{\infty} \frac{1}{16 ^ i (8 i + k) }
\end{equation}
By using the notation  $\left[f(x)\right]_0^{y} = f(y) - f(0)$ and the laws of integration, we get
\begin{equation}\label{eqn:SK1}\label{eqn:SK2}
S_k =  \sqrt{2} ^ k \sum_{i=0}^{\infty}\left[\frac {x ^ {k + 8i}}{8i+k}\right]_0^{\frac{1}{\sqrt{2}}}
=  \sqrt{2} ^ k \sum_{i=0}^{\infty} \int_{0}^{\frac{1}{\sqrt{2}}} x ^ {k - 1 + 8i} \,{\rm d}x
\end{equation}
Thanks to uniform convergence, the series and the integral can be exchanged and
we can then factor out \(x^{k-1}\) and recognize a geometric series in \(x^8\).
\begin{equation}\label{eqn:SK3}\label{eqn:SK4}
S_k = \sqrt{2} ^ k\int_{0}^{\frac{1}{\sqrt{2}}} \sum_{i=0}^{\infty} x ^ {k - 1 + 8i} \,{\rm d}x =
 \sqrt{2} ^ k\int_{0}^{\frac{1}{\sqrt{2}}} \frac {x ^ {k -1}}{1 - x ^8} \,{\rm d}x
\end{equation}
Now replacing the $S_k$ values  in the right hand side of (\ref{eqn:plouffe}),  we get:
\begin{equation}\label{eqn:S0}
S = 4 S_1 -2S_4 -S_5 -S_6 =  \int_{0}^{\frac{1}{\sqrt{2}}} \frac{4\sqrt{2} - 8 x^3 -4\sqrt{2}x^4 -8x^5} { 1 - x^8} \,{\rm d}x
\end{equation}
Then, with the variable change \(y = \sqrt{2} x\) and algebraic calculations
on the integrand
\begin{equation}\label{eqn:S1}
S
=\int_{0}^1 \frac {4 -4y}{y^2 -2y +2} + \frac{4}{1 + (y -1)^2} + 4\frac{y}{y^2 -2}\,{\rm d}y
\end{equation}
We recognize here the respective derivatives of $-2\ln(y^2 -2y +2)$, $4 \arctan(y-1)$ and $2\ln(2-y^2)$.  Most of these functions have null or compensating
values at the bounds of the integral, leaving only one interesting term:
\begin{eqnarray*}
S  &=& \left[-2\ln(y^2 -2y +2)+4 \arctan(y-1)+2\ln(2-y^2)\right]_0^1\\
&=& -4 \arctan(-1) = \pi
\end{eqnarray*}

\subsubsection{The formalization of the proof}

The current version of our formal proof, compatible with Coq version 8.5 and
8.6 \cite{coq}
is available on the world-wide web~\cite{PlouffeAGMSources}.
To formalize this proof, we use  the Coquelicot library intensively. 
This library deals with series, power series, integrals and provides some theorems linking these notions that we need for our proof.
In Coquelicot, series (named {\tt Series})
are defined as in standard mathematics as  the sum of the terms of an infinite sequence (of type $ nat\rightarrow R$ in our case)
and power series ({\tt PSeries}) are the series of terms of the form $a_n x^n$.
The beginning of the formalisation follows the proof (steps ~(\ref{eqn:SK}) to (\ref{eqn:SK2})).
Then, one of the key arguments of the proof  is the exchange of the integral sign  and the series allowing the transition  from equation~(\ref{eqn:SK2}) to equation~(\ref{eqn:SK3}).
The corresponding  theorem provided by Coquelicot is the following:
\begin{verbatim}
Lemma RInt_PSeries (a : nat -> R) (x : R) :
   Rbar_lt (Rabs x) (CV_radius a) ->
   RInt (PSeries a) 0 x = PSeries (PS_Int a) x.
\end{verbatim}
where {\tt (PSeries (PS\_Int} \(a\){\tt{}))}  is the series whose (n+1)-th  term is $ \frac{a_n}{n + 1} x^{n+1}$ coming from the equality:
$\int_0^x a_n x^n = \left[ \frac{a_n}{n + 1} x^{n+1} \right]_0^x$.
We use this lemma as a rewriting rule from right to left.

Note that the {\tt RInt\_PSeries} theorem assumes that the  integrated function is   a power series (not a simple series), that is, a series whose terms have the form  $a_i x^i$.
In our case, the term of the series is $x^{k-1+8i}$, that is $x^{k -1} x^{8i}$.
To transform it into an equivalent power series we have first to transform  the series  $\sum_i x^{8i}$  into a  power series.
For that purpose,  we  define the {\tt hole} function.
\begin{verbatim}
Definition hole (n : nat) (a : nat -> R) (i : nat) :=
  if n mod k =? 0  then a (i / n)  else 0.
\end{verbatim}
and prove the equality given in the following lemma.
\begin{verbatim}
Lemma fill_holes k a x : 
  k <> 0 -> ex_pseries a (x ^ k) ->
  PSeries (hole k a) x = Series (fun n => a n * x ^ (k * n)).
\end{verbatim}
The premise written in the second line of {\tt fill\_holes}
expresses that the series $\sum_i a_i (x^k)^i$ converges.
This equality expresses that the series of term $ a_i (x^k)^i$ is equivalent to the power series which  terms are $a_{n/k}$ when n is a multiple of $k$ and 0 otherwise.

Then by combining {\tt fill\_holes}  with the Coquelicot function {\tt (PS\_incr\_n a n)}, that shifts the coefficients of the series $ \sum_{i=0}^{\infty} a_i x^{n+i}$ to transform it into
$\sum_{i = 0}^{i = n-1} 0.x^{i} + \sum_{i = n}^{\infty} a_{i - n} x^i$ that is a power series, 
we prove the {\tt PSeries\_hole} lemma.
\begin{verbatim}
Lemma PSeries_hole x a d k :
  0 <= x < 1 ->
  Series (fun i : nat => a * x ^ (d + S n * i)) =
  PSeries (PS_incr_n (hole (S k) (fun _ : nat => a)) n) x
\end{verbatim}

\noindent Moreover, the {\tt RInt\_PSeries} theorem contains the
hypothesis
that the absolute value of the upper bound of the integral, that is $|x|$, 
is less than the radius of convergence of the power series associated to $a$.
This is proved in the following lemma.
\begin{verbatim}
Lemma PS_cv x a :
  (forall n : nat, 0 <= a n <= 1) ->
  0 <= x -> x < 1 -> Rbar_lt (Rabs x) (CV_radius a)
\end{verbatim} 
It should be noted that in our case $a_n$ is either $1$ or $0$ and the
hypothesis \verb+forall n : nat, 0 <= a n <= 1+ is easily satisfied.

In summary, the first part of the proof is formalized by the {\tt Sk\_Rint} lemma:
\begin{verbatim}
Lemma Sk_Rint k (a := fun i => / (16 ^ i * (8 * i + k))) :  
  0 < k ->
  Series a = 
   sqrt 2 ^ k  *
       RInt (fun x => x ^ (n - 1) / (1 - x ^ 8)) 0 (/ sqrt 2).
\end{verbatim}
that computes  the value of $S_k$ given by (\ref{eqn:SK3}) from the definition (\ref{eqn:SK}) of {\tt Sk}.

The remaining of the formalized proof  follows closely the mathematical proof described in the previous section.
We first perform an integration by substitution (starting from equation~(\ref{eqn:S0})), replacing the variable $x$ by $\sqrt {2} x$, 
by rewriting (from right to left) with the \verb+RInt_comp_lin+ Coquelicot lemma. 
\begin{verbatim}
Lemma RInt_comp_lin f u v a b :
  RInt (fun y : R => u * f (u * y + v)) a b =
  RInt f (u * a + v) (u * b + v)
\end{verbatim}
This  lemma assumes that the substitution function is a linear function, which is the case here.

\noindent Then we decompose {\tt S} into three parts (by
computation) to obtain equation~(\ref{eqn:S1}), actually decomposed into
three integrals that are computed in lemmas
{\tt RInt\_Spart1}, {\tt RInt\_Spart2}, and {\tt RInt\_Spart3} respectively.
For instance:
\begin{verbatim}
Lemma RInt_Spart3 : 
  RInt (fun x => (4 * x) / (x ^ 2 - 2)) 0 1 = 2 * (ln 1 - ln 2).
\end{verbatim}
Finally, we obtain the final result, based on the equality
\(\arctan 1 = \frac{\pi}{4}\).

\subsection{ Computing the nth  decimal of $\pi$ using the Plouffe formula}
 \label{sec:plouffe_C}
 
We now describe how the formula  (\ref{eqn:plouffe}) can be used to compute a 
 particular decimal of $\pi$ effectively. This formula is a summation of four terms where
 each term has the form  ${1}/{16 ^ i (8 i + k)}$ for some $k$. Digits are then expressed in
hexadecimal (base 16). Natural numbers strictly less than $2^p$ are used to simulate
a modular arithmetic with $p$ bits, where $p$ is the precision of computation.
We first explain how the computation of $S_k= \sum_i{{1}/{16 ^ i (8 i + k)}}$ for a given $k$ is performed. Then, we describe
how the four computations are combined to get the final digit.

We want to get the digit at position \(d\).  The first operation is to scale
the sum $S_k$ by a factor $m=16^{d -1}\, 2^p$ to be able to use integer
arithmetic.  In what follows, we need that \(p\) is greater than four.
 If we consider $\lfloor mS_k\rfloor$ (the integer part of
$mS_k$), the
digit we are looking for is composed of its bits $p$, $p-1$, $p-2$,
$p-3$ that can be computed using basic integer operations: $(\lfloor
mS_k\rfloor \texttt{\,mod\,} 2^p) / 2 ^{p - 4}$. Using integer
arithmetic, we are going to compute an approximation of $\lfloor
mS_k\rfloor \texttt{\,mod\,} 2^p$ by splitting the sum into three parts
\begin{equation}\label{eqn:Cplouf1}
mS_k =  \sum_{0 \le i < d} \frac{m}{16^i (8 i + k)} +
      \sum_{d \le i < d + p /4} \frac{m}{16^i(8 i + k)} + 
      \sum_{d + p / 4 \le i} \frac{m}{16^i (8 i + k)}
\end{equation}
In the first part, the inner term can be rewritten as $\frac{2^p 16^{d- 1 -i}}{8 i + k}$
where both divisor and dividend are natural numbers.  The division
can be performed in several stages.  To understand this, it is worth
comparing the fractional and integer part of \(\frac{16 ^
  {d-1-i}}{8i+k}\) with the bits of \(\frac{2^p 16 ^ {d-1-i}}{8i+k}\).

For illustration, let us consider the case where \(i = 0\), \(k=3\),
\(p=4\), and \(d = 2\).  The number we wish to compute is
\[\frac{2^416^{2-1}}{3}\]
and we only need to know the first 4 bits, that is we need to know
this number modulo \(2^4\). The ratio is \(85.33\overline{3}\), and modulo 16
this is 5.  Now, we can look at the number \(2^4 \frac{16}{3}\).
If we note \(q\) and \(r\) the quotient and the remainder of the
division on the left (when viewed as an integer division), we have
\[2^4 \frac{16}{3} = 2^4 q + \frac{2 ^ 4 r}{3}\]
Since we eventually want to take this number modulo \(2^4\),
the left part of the sum, \(2^4 q\), does not impact the result and
we only need to compute \(r\), in other words \(16 \texttt{\,mod\,}
3\).  In our illustration case, we have \(16 \texttt{\,mod\,} 3 = 1\)
and \(\frac{2 ^ 4 \times 1}{3} = 5.333\), so we do recover the right
4 bits.  Also, because we are only interested in bits that are part
of the integral part of the result, we can use integer division to
perform the last operation.

These computations are performed in the following Coq function, that
progresses by modifying a state datatype containing the current index
and the current sum.  In this function, we also take care of keeping
the sum under \(2^p\), because we are only concerned with this sum
modulo \(2^p\).
\begin{verbatim}
Inductive NstateF := NStateF (i : nat) (res : nat). 
\end{verbatim}
Doing an iteration is performed by
\begin{verbatim}
Definition NiterF k (st : NstateF) :=
  let (i, res) := st in
  let r := 8 * i + k in
  let res := res + (2 ^ p * (16 ^ (d - 1 - i) mod r)) / r in
  let res := if res < 2 ^ p then res else res - 2 ^ p in
  NStateF (i + 1) res.
\end{verbatim}
The summation is performed by  $d$ iterations:
\begin{verbatim}
Definition NiterL k := iter d  (NiterF k)  (NStateF 0 0).
\end{verbatim}
The result of \texttt{NiterL} is a natural number. What we need
to prove is that it is a modular result and it is not so far from
the real value. As we have turned an exact division into a division over natural numbers, the error is at most 1.
After $d$ iterations, it is at most $d$. This is stated by the following lemma.
\begin{verbatim}
Lemma sumLE k (f := fun i => ((16 ^ d / 16) * 2 ^ p) / 
                              (16 ^ i * (8 * i + k))) :
  0 < k ->
  let (_, res) := NiterL p d k in
  exists u : nat, 0 <= sum_f_R0 f (d - 1) - res - u * 2 ^ p < d.
\end{verbatim}
where \texttt{sum\_f\_R0 f n} represents the summation $f(0) + f(1) + \dots +  f(n)$. 

Let us now turn our attention to the second part of the iteration of
formula~(\ref{eqn:Cplouf1}).
\[\sum_{d \le i < d + p /4} \frac{m}{16^i(8 i + k)} = \sum_{d \le i < d + p /4}  \frac{2^p16^{d-1-i}}{8i+k}.\]

All the terms of this sum are less than $2^p$. As terms get smaller
by a factor of at least 16, we consider only $p/4$ terms.
We first build a datatype that contains the current index, the current shift and the current result:
\begin{verbatim}
Inductive NstateG := NStateG (i : nat) (s : nat) (res : nat). 
\end{verbatim}
We then define what is a step:
\begin{verbatim}
Definition NiterG k (st : NstateG) :=
  let (i, s, res) := st in
  let r := 8 * i + k in
  let res := res + (s / r) in
  NStateG (i + 1) (s / 16) res.
\end{verbatim}
and we iterate $p/4$ times:
\begin{verbatim}
Definition NiterR k :=
  iter (p / 4)  (NiterG k) (NStateG d (2 ^ (p - 4)) 0).
\end{verbatim}
Here we do not need any modulo since the result fits in $p$ bits
and as the contribution of each iteration makes an error of at most one unit
with the division by $r$, the total error is then bounded by $p/4$. This is
stated by the following lemma.
\begin{verbatim}
Lemma sumRE k (f := fun i => 
                      ((16 ^ d / 16) * 2 ^ p) / 
                       (16 ^ (d + i) * (8 * (d + i) + k))) :
  0 < k -> 0 < p / 4 ->
  let (_, _, s1) := NiterR k in
  0 <= sum_f_R0 f (p / 4 - 1) - s1 < p / 4.
\end{verbatim}

The last summation is even simpler. We do not need to perform any computation.
all the terms are  smaller than 1 and quickly decreasing. It is then easy to prove
that this summation is strictly smaller than 1.

Adding the two computations, we get our approximation.
\begin{verbatim}
Definition NsumV k :=
  let (_, res1) := NiterL k in
  let (_, _, res2) := NiterR k in res1 + res2. 
\end{verbatim}
We know that it is an under approximation and the error is less than $d + p / 4 + 1$.

We are now ready to define our function that extracts the digit:
\begin{verbatim}
Definition NpiDigit :=
  let delta := d + p / 4 + 1 in
  if (3 < p) then
    if 8 * delta < 2 ^ (p - 4) then
      let Y := 4 * (NsumV 1) + 
             (9 * 2^ p -
             (2 * NsumV 4 + NsumV 5 + NsumV 6 + 4 * delta)) in
      let v1 := (Y + 8 * delta) mod 2 ^ p / 2 ^ (p - 4) in
      let v2 := Y mod 2 ^ p / 2 ^ (p - 4) in
      if v1 = v2 then Some v2 else None
    else None
  else None.
\end{verbatim}
This deserves a few comments.
In this function, the variable  \texttt{delta} represents the error that is
   done by one application of \texttt{NsumV}.  When adding the
   different sums, we are then going to make an overall error of
 \texttt{8 * delta}.  Moreover, we know that \texttt{NsumV} is an
under approximation.  The variable {\tt Y} computes an under approximation
of the result: for those sums that appear negatively, the under
approximation is obtained adding \texttt{delta} to the sum before taking
the opposite.  This explains the fragment \texttt{... + 4 * delta}
that appears on the seventh line.  Each of the sums obtained by
{\tt NsumV} actually is a natural number \(s\) smaller than \(2 ^ p\), when
it is multiplied by a negative coefficient, this should
be represented by \(2 ^ p - s\).  Accumulating all the compensating
instances of \(2 ^ p\) leads to the fragment \verb"9 * 2 ^ p - ..."
that appears on the sixth line.

After all these computations, \texttt{Y +  8 * delta} is an
over approximation. If both 
  \texttt{Y} and  $\texttt{Y + 8 * delta}$ give the same digit,
we are sure that this digit is valid. 

The correctness of the \texttt{NpiDigit} function is proved with
respect to the definition of what is the digit at place \(d\)
in base $b$ of a real number $r$, i.e.
we take the integer part of $r  b^d$ and we take the modulo $b$:
\begin{verbatim}
Definition Rdigit (b : nat) (d : nat) (r : R) := 
  (Int_part ((Rabs r) * (b ^ d))) mod b.
\end{verbatim}
The correctness is simply stated as
\begin{verbatim}
Lemma NpiDigit_correct k : 
  NpiDigit = Some k -> Rdigit 16 d PI = k.
\end{verbatim}
Note that this is a partial correctness statement. A program that always
returns \texttt{None} also satisfies this statement. If we look at
the actual program, it is clear that one can precompute a $p$ that
fulfills the first two tests, the equality test is another story.  A
long sequence of 0 (or F) may require a very high precision.

This program is executable but almost useless since it is based
on a Peano representation of the natural numbers.  Our next step was
to derive an equivalent program using a more efficient representation
of natural numbers, provided by the type \texttt{BigN}~\cite{GregoireTheryIJCAR2006}.
This code also receives some optimizations to implement faster
operations of multiplications and divisions by powers of 2 and
fast modular exponentiations.

Computing within Coq that 2 is the
millionth decimal in hexadecimal of $\pi$ with a precision
of 28 bits (27 are required for the first two tests and 28 for the equality
test) takes less than 2 minutes.
In order to reach the billionth decimal, we implement a very
naive parallelization for a machine with at least four cores:
each sum is computed on a different core generating a theorem then the final
result is computed using these four theorems.
With this technique, we get the
millionth decimal, 2, in 25 seconds and the billionth decimal, 8, in 19
hours. Note that we could further parallelize inside the individual sums
 to compute partial sums and then use Coq theorems to glue them together.

\section{Algorithms to compute \(\pi\) based on arithmetic geometric means}
\label{sec:agm}
In principle, all the mathematics that we had to describe formally in
our study of arithmetic geometric means and the number \(\pi\) are available
from the mathematical litterature, essentially from the monograph by
J.~M.~Borwein and P.~B.~Borwein~\cite{BorweinAGM} and the initial papers by
R.~Brent~\cite{Brent76}, E.~Salamin~\cite{Salamin76}.  However, we had
difficulties using these sources as a reference, because they rely on an
extensive mathematical culture from the reader.  As a result, we were actually
guided by a variety of sources on the world-wide web, including an
exam for the selection of French high-school mathematical
teachers~\cite{CapesAGM95}.  It feels useful to repeat these mathematical
facts in a first section, hoping that they are exposed at a sufficiently
elementary level to be understood by a wider audience.  However, some
details may still be missing from this exposition and they can be recovered
from the formal development itself.

This section describes two algorithms, but their mathematical justification
has a lot in common.  The first algorithm that we present came to us as the
object of an exam for high-school teachers~\cite{CapesAGM95}, but in reality
this algorithm is neither the first one to have been designed by
mathematicians, nor the most efficient of the two.  However, it is interesting
that it brings us good tools to help proving the second one, which is
actually more traditional (that second algorithm dates from 1976
\cite{Brent76,Salamin76}, and it is
the one implemented in the {\tt mpfr} library \cite{mpfr}) and more efficient
(we shall see that it requires much less divisions).

In a second part of our study, we concentrate on the accumulation of errors
during the computations and show that we can also prove bounds on this.  This
part of our study is more original, as it is almost never covered
in the mathematical litterature, however it re-uses most of the results
we exposed in a previous article~\cite{Bertot:CPP15}.
\subsection{Mathematical basics for arithmetic geometric means}
\label{sec:math-agm}
Here we enumerate a large collection of steps that make it possible to go
from the basic notion of arithmetic-geometric means to the computation of
a value of \(\pi\), together with estimates of the quality of approximations.

This is a long section, consisting of many simple facts, but some of the
detailed computations are left untold.  Explanations given between the
formulas should be helpful for the reader to recover most of the steps.
However, missing information can be found directly in the actual formal
development~\cite{PlouffeAGMSources}.
\paragraph{The arithmetic-geometric process.}
As already explained in section~\ref{intro}, the arithmetic{\hskip0pt}-geometric mean of
two numbers \(a\) and \(b\) is obtained by defining sequences \(a_n\) and
\(b_n\) such that \(a_0 = a\), \(b_0 = b\) and
\[a_{n+1} = \frac{a_n + b_n}{2} \qquad b_{n+1} = \sqrt{a_n b_n}\]

A few tests using high precision calculators show that the two sequences
\(a_n\) and \(b_n\) converge rapidly to a common value \(M(a,b)\), with
the number of common digits doubling at each iteration.  The sequence
 \(a_n\) provides
over approximations and the sequence \(b_n\) under approximations.
Here is an example
computation (for each line, we stopped printing values at the first differing
digit between \(a_n\) and \(b_n\)).
\begin{center}
\begin{tabular}{|r|r|r|}
\hline
&\(a\)&\(b\)\\
\hline
0&1&0.5\\
1&0.75&0.70\dots\\
2&0.7285\dots&0.7282\dots\\
3&0.72839552\dots&0.72839550\dots\\
4&0.7283955155234534\dots&0.7283955155234533\dots\\
\hline
\end{tabular}
\end{center}
The function \(M(a,b)\) also benefits from a scalar multiplication property:
\begin{equation}
M(ka, kb) = k M(a, b)\qquad M(a,b) = a M(1, \frac{b}{a})
\end{equation}
For the sake of computing approximations of \(\pi\), we will mostly be
interested in the sequences \(a_n\) and \(b_n\) stemming from \(a_0=1\)
and \(b_0 = \frac{1}{\sqrt{2}}\).
\paragraph{Elliptic integrals.}
We will be interested in
\emph{complete elliptic integrals of the first kind}, noted \(K(k)\).
The usual definition of these integrals has the following form
\begin{equation}K(k) = \int_0^{\frac{\pi}{2}} \frac{{\rm d}\theta}
  {\sqrt{1 - k^2 \sin^2 \theta}}
\end{equation}
But it can be proved that the following equality holds, when setting
 \(a = 1\) and \(b = \sqrt{1- k^2}\), and using a change of variable
(we only use the form \(I(a,b)\)):
\begin{equation}\label{eqn:elliptic-integral-def}
K(k) = I(a, b) = \int_0^{+\infty} \frac{{\rm d}t}
{\sqrt{(a ^ 2 + t ^ 2) (b ^ 2 + t ^ 2)}}
\end{equation}
Note that the integrand in \(I\) is symmetric, so that \(I(a,b)\) is also
half of the integral with infinities as bounds.
With the change of variables \(s = \frac{1}{2}(x - \frac{ab}{x})\), then
reasoning by induction and taking the limit, we
also have the following equalities
\begin{equation}\label{eqn:elliptic-agm-step}
I(a,b) = I(\frac{a + b}{2}, \sqrt{ab})=I(a_n,b_n)=I(M(a,b),M(a,b))=\frac{\pi}{2M(a,b)}.
\end{equation}
\paragraph{Equivalence when \(x \rightarrow 0\) and derivatives.}
Another interesting property for elliptic integrals of the first kind can
be obtained by the variable change \(u = \frac{x}{t}\) on the
integral on the right-hand side of equation (\ref{eqn:elliptic-integral-def}).
\begin{equation}\label{eqn:elliptic-sqrt-bound}
I(1, x) = 2 \int_0^{\sqrt{x}}\frac{{\rm d}t}
{\sqrt{(1 + t ^ 2)(x + t ^ 2)}}
\end{equation}
Studying this integral when \(x\) tends to \(0\) gives the
equivalences for \(I\) and \(M\):
\begin{equation}\label{eqn:Iequiv}\label{eqn:Mequiv}
I(1,x) \sim 2 \ln(\frac{1}{\sqrt{x}})\quad
M(1, x) \sim \frac{-\pi}{2 \ln x} \qquad\hbox{when}~x\rightarrow 0^+.
\end{equation}

For the rest of this section, we will assume that \(x\) is a value in
the open interval \((0, 1)\) and that \(a_0 = 1\) and \(b_0 = x\).
Coming back to the sequences \(a_n\) and \(b_n\), the following property
can be established.
\begin{equation}
M\left(a_{n+1}, \sqrt{a_{n+1}^2 - b_{n+1}^2}\right) = \frac{1}{2} M\left(a_n, \sqrt{a_n^2 - b_n^2}\right)
\end{equation}
We can repeat \(n\) times and use the fact that \(a_0^2- b_0^2 = 1 - x^2\).
\begin{equation}
2 ^ n M \left(a_n, \sqrt{a_n ^ 2 - b_n ^ 2}\right) =
 2 ^ n a_n M\left(1, \frac{\sqrt{a_n ^ 2 - b_n ^ 2}}{a_n}\right)
=
M \left(1, \sqrt{1 - x ^ 2}\right)
\end{equation}
Still under the assumption of \(a_0 = 1\) and \(b_0 = x\),
we define \(k_n\)  as follows:
\begin{equation}
k_n(x) = \frac{\ln \left(\frac{a_n}{\sqrt{a_n ^ 2 - b_n ^ 2}}\right)}{2 ^ n}
\end{equation}
Through separate calculation, involving Equation~(\ref{eqn:Mequiv}) and
the definition of \(k\), we establish the following properties.
\begin{equation}\label{eqn:first_derivative_k}
\lim_{n \rightarrow \infty} k_n (x) = \frac{\pi}{2} \frac{M(1, x)}{M(1,\sqrt{1-x^2})}
\qquad
k_n' = \frac{b_n ^ 2}{x (1 - x ^ 2)}
\end{equation}
These derivatives converge uniformly to their
limit.  Moreover, the sequence of derivatives
of \(a_n\) is growing and converges uniformly.  This guarantees that
\(x \mapsto M(1,x)\) is also differentiable and and its derivative is
the limit of the derivatives
of \(a_n\).  We can then obtain the following two equations, the second
is our main central formula.
\begin{equation}\label{eqn:main_derivative}
\label{eqn:pi_from_agm_and_derivative}
\left(\frac{\pi}{2} \frac{M(1, x)}{M(1, \sqrt{1 - x ^ 2})}\right)'
 =
\frac{M(1, x) ^ 2}{x (1 - x ^ 2)}
\qquad
\pi = 2 \sqrt{2} \frac{M(1, \frac{1}{\sqrt{2}})^3}
                      {\displaystyle
                           \left(M(1, x)\right)'(\frac{1}{\sqrt{2}})}
\end{equation}

We define the functions \(y_n = \frac{a_n}{b_n}\) and \(z_n = \frac{b'_n}{a'_n}\).  These sequence satisfy
\begin{equation}
y_0 = \frac{1}{x} \qquad y_{n+1} = \frac{1 + y_n}{2\sqrt{y_n}} \qquad
z_1 = \frac{1}{\sqrt{x}} \qquad z_{n+1} = \frac{1 + z_n y_n}{(1 + z_n) \sqrt{y_n}}
\end{equation}
and the following important chain of comparisons.
\begin{equation}\label{eqn:chain_y_z_y}
y_{n+1} \leq z_{n+1} \leq \sqrt{y_n}
\end{equation}
\paragraph{Computing with \(y_n\) and \(z_n\) (the Borwein algorithm).}  The
 first algorithm we will present, proposed
by J.~M.~Borwein and P.~B.Borwein, consists in approximating \(M\)
using the sequences \(y_n\) and \(z_n\).  
The value
\(M(1, x)^3\) is approximated using \(a_n b_n^2\) and \((M(1,x))'\) using
\(a_n'\), all values being taken in \(\frac{1}{\sqrt{2}}\).

From the definition, we can easily derive the following
properties:
\begin{equation}\label{eqn:y_step_z_step}
{1 + y_n} = 2 \frac{a_{n+1} b_{n+1}^2}{a_{n} b_{n}^2}
\qquad
{1 + z_n} = 2 \frac{a_{n+1}'}
{a_n'}
\end{equation}
Repeating the products, we get the following definition of a sequence
\(\pi_n\) and the proof of its limit:
\begin{equation}
\pi_0 = (2 + \sqrt{2}) \qquad \pi_n = \pi_0\prod_{i=1}^{n}
  \frac{1 + y_i}{1 + z_i}
\qquad \lim_{n\rightarrow\infty} \pi_n = \pi
\end{equation}
\paragraph{Convergence speed.} For an arbitrary \(x\) in the open interval
\((0, 1)\), using a Taylor expansion of the function
\(y \mapsto \frac{1 + y}{2\sqrt{y}}\) of order two, and then reasoning by
induction, we get the following results:
\begin{equation}
y_{n+1}(x) - 1 \leq \frac{(y_n(x) - 1) ^ 2}{8}
\qquad
y_{n + 1}(x) \leq 8 \left(\frac{(y_1(x) - 1)}{8}\right)^{2 ^ n}
\end{equation}
For \(x = \frac{1}{\sqrt{2}}\), we obtain the
following bound:
\begin{equation}\label{bound_y}
y_{n+1}(\frac{1}{\sqrt{2}}) - 1 \leq 8 \times 531^{-2^n}
\end{equation}
Using the comparisons of line~(\ref{eqn:chain_y_z_y}) and then reasoning by
induction we obtain our final error estimate:
\begin{equation}\label{eqn:bound_pi}
0 \leq \pi_{p+1} - \pi \leq \pi_{p+1} \left(y_{p+1}(\frac{1}{\sqrt{2}}) - 1 \right)
\leq 4 \pi_0 531 ^{-2 ^ p}
\end{equation}
\paragraph{Computing one million decimals.}
The first element of the sequence \(\pi_n\) that is close to
\(\pi\) with an error smaller than \(10 ^ {10 ^ 6}\) is obtained for
\(n\) satisfying the following comparison.
\begin{equation}\label{bound-agmpi}
n \geq
\frac{\ln\left(\frac{10 ^ 6  \ln 10 - \ln(4  \pi_0)}{\ln 531}\right)}
{\ln 2} \sim 18.5
\end{equation}
For one million hexadecimals, \(n\) only needs to be larger than \(18.75\).

\paragraph{Computing with an infinite sum (the Brent-Salamin algorithm).}
The formula described in this section probably appears in Gauss' work and
is repeated by King \cite{King1924}.  It was published and clarified for
implementation on modern computers by
Brent~\cite{Brent76} and Salamin~\cite{Salamin76}.  A good account of
the historical aspects is given by Almkvist and Berndt \cite{AlmkvistBerndt88}.
Our presentation relies on a mathematical exposition given by
Gourevitch~\cite{Gourevitch99}.

In the variant proposed by Brent and Salamin, we compute the right-hand
side of the main central formula by computing \(a_n^2\) and the ratio
\(\frac{b_n'}{b_n}\).  We first introduce a third function \(c_n\).
\begin{equation}\label{c_step}
c_n = \frac{1}{2}(a_{n-1} - b_{n-1})
\end{equation}
The derivative of function \(k_n\) can be expressed with
\(c_n\) and after combination with equation~\ref{eqn:first_derivative_k},
this gives a formula for the derivative of \(\frac{a_n}{b_n}\) at
\(\frac{1}{\sqrt{2}}\).
\begin{equation}\label{eqn:derivative_k_extra}
\label{eqn:derivative_ratio_a_b_direct_value}
\left(\frac{a_n}{b_n}\right)'(\frac{1}{\sqrt{2}}) =
\frac{-2^{n+1} \sqrt{2} \, a_n c_n^2}
{b_n}
\end{equation}
The derivative of this ratio can be compared to the difference
of the ratio of \(b_n'\) over \(b_n\) at two successive
indices, which can be repeated \(n\) times.
\begin{equation}\label{eqn:diff_ratio_b'_b}\label{salamin_sum}
\frac{b_{n+1}'}{b_{n+1}}
- \frac{b_n'}{b_n}
=\frac{b_n}{2 a_n}\left(\frac{a_n}{b_n}\right)'
\qquad
\frac{b_{n+1}'}{b_{n+1}} =
\frac{b_1'}{b_1} -\sqrt{2} \sum_{k = 1}^{n-1} 2 ^ k c_n^2
\end{equation}
We can then use equations~(\ref{c_step})
 and~(\ref{eqn:pi_from_agm_and_derivative}), where
\(M^3(1,\frac{1}{\sqrt{2}})\)
is the limit of \(a_n^2 b_n\) to obtain the final definition
and limit.
\begin{equation}\label{pi_salamin_def}\label{salamin_complete_formula}
\pi'_n = \frac{4\, a_n^2}
{1 - \sum_{k=1}^{n-1} 2 ^ {k-1}(a_{k - 1} - b_{k - 1}) ^ 2}
\qquad
\pi = \lim_{n\rightarrow \infty} \pi'_n
\end{equation}
\paragraph{Speed of convergence.}  We can link the Brent-Salamin algorithm
with the Borwein algorithm in the following manner:
\begin{equation}
\pi'_n = 2\sqrt{2}\,\frac{a_n^2 b_n}{b_n'} =
 2\sqrt{2}\, \frac{y_n}{z_n} \frac{a_n b_n^2}{a_n'} =
\frac{y_n}{z_n} \pi_n
\end{equation}
Combining bounds~(\ref{eqn:chain_y_z_y}), (\ref{bound_y}),
and~(\ref{eqn:bound_pi}) we obtain this first approximation.
\begin{equation}\label{eqn:first_bound_pi'}
| \pi'_{n+1} - \pi| \leq  68 \times 531 ^ {-2^{n-1}}
\end{equation}
This first approximation is too coarse, as it gives the impression that
\(\pi'_{n+1}\) is needed when \(\pi_{n}\) is enough (the exponent of
2 in bound~(\ref{eqn:first_bound_pi'}) is \(n-1\) while
it is \(n\) in bound~(\ref{eqn:bound_pi})).  We can
make it better by noting that the difference between \(\pi\)
and \(\pi_{n+2}\) is \(O(531^{-2^n})\) and the difference between \(\pi_{n+2}\)
and \(\pi_{n+1}\) is significantly smaller than \(531^{-2^{n-1}}\), while
not being \(O(531^{-2^n})\).
\begin{equation}\label{eqn:bound_pi'}
|\pi'_{n+1} -\pi | \leq (132 + 384 \times 2^n) \times 531 ^{-2^n}
\end{equation}
For one million decimals of \(\pi\), we can still use \(n=19\).

Each algorithm computes \(n\) square roots to compute \(\pi_n\) or
\(\pi'_n\).  However, the first one uses \(3n\) division to obtain
value \(\pi_n\), while the second one only performs divisions by
\(2\), which are less costly, and a single full division at the end of
the computation to compute \(\pi_n'\).  In our experiments computing
these algorithms inside Coq, the second one is twice as fast.

\subsection{Formalization issues for arithmetic geometric means}
In this section, we describe the parts of our development where we
had to proceed differently from the mathematical exposition in
section~\ref{sec:math-agm}.  Many difficulties arose from gaps in
the existing libraries for real analysis.
\paragraph{The arithmetic geometric mean functions.}  For a given
\(a_0 = a\) and \(b_0 = b\), the functions \(a_n\) and \(b_n\)
actually are functions of \(a\) and \(b\) that are defined mutually recursively.
Instead of a mutual recursion between two functions, we chose to
simply describe a function {\tt ag} that takes three arguments and
returns a pair of two arguments.  This can be written in the following
manner:
\begin{verbatim}
Fixpoint ag (a b : R) (n : nat) :=
  match n with
    0 => (a, b)
  | S p => ag ((a + b) / 2) (sqrt (a * b)) p
  end.

\end{verbatim}
This functions takes three arguments, two of which are real numbers,
and the third one is a natural number.  When the natural number
is \(0\), then the result is the pair of the real numbers, thus
expressing that \(a_0 = a\) and \(b_0 = b\).  When the natural
number is the successor of some \(p\), then the two real number
arguments are modified in accordance to the arithmetic-geometric mean
process, and then the \(p\)-th argument of the sequence starting
with these new values is computed.

As an abbreviation we also use the following definitions, for the special
case when the first input is \(1\).
\begin{verbatim}
Definition a_ (n : nat) (x : R) := fst (ag 1 x n).

Definition b_ (n : nat) (x : R) := snd (ag 1 x n).
\end{verbatim}

The function {\tt ag\_step}  seems to perform the
operation in a different order, but in fact we can really show
that \(a_{n+1} = \frac{a_n + b_n}{2}\) and \(b_{n+1} = \sqrt{a_n b_n}\) as
expected, thanks to a proof by induction on \(n\).  This is expressed
with theorems of the following form:
\begin{verbatim}
Lemma a_step n x : a_ (S n) x = (a_ n x + b_ n x) / 2.

Lemma b_step n x : b_ (S n) x = sqrt (a_ n x * b_ n x).
\end{verbatim}
\paragraph{Limits and filters.}
Cartan \cite{Filtres-cartan} proposed in 1937 a general notion that
made it possible to develop notions of limits in a uniform way, whether
they concern limits of continuous function of limits of sequences.  This
notion, known as \emph{filters} is provided in formalized mathematics in
Isabelle \cite{Isabelle-filters}
and more recently in the Coquelicot library \cite{BLM15}.  It is
also present in a
simplified form as {\em convergence nets} in Hol-Light \cite{Harrison98}.

Filters are not real
numbers, but objects designed to represent ways to approach a limit.
There are many kinds of filters, attached to a wide variety of
types, but for our purposes we will mostly be interested in seven
kinds of filters.
\begin{itemize}
\item {\tt eventually} represents the limit towards \(\infty\), but
  only for natural numbers,
\item {\tt locally \(x\)} represents a limit approaching a real number
  \(x\) from any side,
\item {\tt at\_point \(x\)} represents a limit that is actually not a
  limit but an exact value: you approach \(x\) because you are bound
  to be exactly \(x\),
\item {\tt at\_right \(x\)} represents a limit approaching \(x\) from
  the right, that is, only taking values that are greater than \(x\)
  (and not \(x\) itself),
\item {\tt at\_left \(x\)} represents a limit approaching \(x\) from
  the left,
\item {\tt Rbar\_locally p\_infty} describes a limit going to
  \(+\infty\),
\item {\tt Rbar\_locally m\_infty} describes a limit going to
\(-\infty\).
\end{itemize}
There is a general notion called {\tt filterlim \(f\) \(F_1\) \(F_2\)}
to express that the value returned by \(f\) tends to a value described
by the filter \(F_2\) when its input is described by \(F_1\).  For
instance, we constructed formal proofs for the following two theorems:
\begin{verbatim}
Lemma lim_atan_p_infty :
  filterlim atan (Rbar_locally p_infty) (at_left (PI / 2)).

Lemma lim_atan_m_infty :
  filterlim atan (Rbar_locally m_infty) (at_right (-PI / 2)).
\end{verbatim}
In principle, filters make it possible to avoid the usual
\(\varepsilon-\delta\) proofs of topology and analysis, using faster
techniques to relate input and output filters for continuous functions
\cite{Isabelle-filters}.
In practice, for precise proofs like the ones above (which use the
{\tt at\_right} and {\tt at\_left} filters), we still need to revert
to a traditional \(\varepsilon-\delta\) framework.
\paragraph{Improper integrals.}
The Coq standard library of real numbers has been providing proper integrals
for a long time, more precisely \emph{Riemann integrals}.  The Coquelicot
library adds an incomplete treatement of improper integrals on top of this.
For improper integrals the bounds are described as limits rather than
as direct real numbers.  For the needs of this experiment, we need to
be able to cut improper integrals into pieces, perform variable changes,
and compute the improper integral
\begin{equation}
\int_{-\infty}^{+\infty} \frac{{\rm d}t}{1 + t ^ 2} = \pi
\end{equation}
The Coquelicot library provides two predicates to describe improper integrals,
the first one has the form\footnote{the name can be decomposed in {\tt
    is} {\tt R} for Riemann, {\tt Int} for Integral, and {\tt gen} for
generalized.}
\begin{alltt}
is_Rint_gen \(f\) \(B\sb{1}\) \(B\sb{2}\) \(v\)
\end{alltt}
The meaning of this predicate is ``the improper integral of function \(f\)
between bounds \(B_1\) and \(B_2\) converges and has value \(v\)''.  The
second predicate is named \verb+ex_Rint_gen+ and it simply takes the
same first three arguments as \verb+is_Rint_gen+, to express that there
exists a value \(v\) such that \verb+is_Rint_gen+ holds.  The Coquelicot
library does not provide a functional form, but there is a general
tool to construct functions from relations where one argument is
uniquely determined by the others, called {\tt iota} in that library.

Concerning elliptic integrals, as a first step we need to express the
convergence of the improper integral in
equation~(\ref{eqn:elliptic-integral-def}).  For this we need a general
theorem of bounded convergence, which is described formally in
our development, because
it is not provided by the library.  Informally, the statement is
that the improper integral of a positive function is guaranteed to
converge if that function is bounded above by another function that
is known to converge.  Here is the formal statement of this theorem:
\begin{verbatim}
Lemma ex_RInt_gen_bound (g : R -> R) (f : R -> R) F G
  {PF : ProperFilter F} {PG : ProperFilter G} :
  filter_Rlt F G ->
  ex_RInt_gen g F G ->
  filter_prod F G
    (fun p => (forall x, fst p < x < snd p -> 0 <= f x <= g x) /\
       ex_RInt f (fst p) (snd p)) ->
    ex_RInt_gen f F G.
\end{verbatim}
This statement exhibits a concept that we needed to devise, the
concept of comparison between filters on the real line, which we
denote {\tt filter\_Rlt}.  This concept will be described in further
detail in a later section.  Three other lines in this theorem
statement deserve more explanations, the lines starting at
{\tt filter\_prod}.  These lines express that a property must 
ultimately be
satisfied for pairs \(p\) of real numbers whose components
tend simultaneously to the
limits described by the filters {\tt F} and {\tt G}, which here
also serve as bounds for two generalized Riemann integrals.
This property is the conjunction of two facts, first for any argument
between the pair of numbers, the function {\tt f} is
non-negative and  less than or equal to {\tt g} at that argument,
second the function {\tt f} is Riemann-integrable between the pair
of numbers.

Using this theorem of bounded convergence, we can prove that the
function
\[ x \mapsto \frac{1}{\sqrt{(x ^ 2 + a ^ 2)(x ^ 2 + b ^ 2)}}\]
is integrable between \(-\infty\) and \(+\infty\) as soon as
both \(a\) and \(b\) are positive, using the function
\[ x \mapsto \frac{1}{m^2 (\left(\frac{x}{m}\right) ^ 2 + 1)} \]
as the bounding function, where \(m = min(a,b)\), and then proving
that this one is integrable
by showing that its integral is related to the arctangent function.

Having proved the integrability, we then define a function that
returns the following integral value:
\[\int_{-\infty}^{+\infty} \frac{{\rm d} x}
    {\sqrt{(x ^ 2 + a ^ 2)(x ^ 2 + b ^ 2)}} \]
The definition is done in the following two steps:
\begin{verbatim}
Definition ellf (a b : R) x :=
   /sqrt ((x ^ 2 + a ^ 2) * (x ^ 2 + b ^ 2)).

Definition ell (a b : R) :=
  iota (fun v => is_RInt_gen (ellf a b)
             (Rbar_locally m_infty) (Rbar_locally p_infty) v).
\end{verbatim}
The value of {\tt ell \(a\) \(b\)} is properly defined when
\(a\) and \(b\) are positive.  This is expressed with the following
theorems, and will be guaranteed in all other theorems where
{\tt ell} occurs.
\begin{verbatim}
Lemma is_RInt_gen_ell a b : 0 < a -> 0 < b ->
  is_RInt_gen (ellf a b)
     (Rbar_locally m_infty) (Rbar_locally p_infty) (ell a b).

Lemma ell_unique a b v : 0 < a -> 0 < b ->
  is_RInt_gen (ellf a b)
     (Rbar_locally m_infty) (Rbar_locally p_infty) v ->
  v = ell a b.
\end{verbatim}
\paragraph{An order on filters.}  On several occasions, we need to
express that the bounds of improper integrals follow the natural order
on the real line.  However, these bounds may refer to no real point.
For instance, there is no real number that corresponds to the limit
\(0+\), but it is still clear that this limit represents a place on
the real line which is smaller than \(1\) or \(+\infty\).  This kind of
comparison is necessary in the statement of \verb+ex_RInt_gen_bound+,
as stated above, because the comparison between functions would be
vacuously true when the bounds of the interval are interchanged.

We decided to introduce a new concept, written {\tt filter\_Rlt \(F\)
  \(G\)} to express that when \(x\) tends to \(F\) and \(y\) tends to
\(G\), we know that ultimately \(x < y\).  To be more precise about
the definition of {\tt filter\_Rlt}, we need to know more about the
nature of filters.

Filters simply are sets of sets.  Every filter contains
 the complete set of
elements of the type being considered, it is stable by
intersection, and it is stable by the operation of taking a
superset.  Moreover, when a filter does not contain the empty set, it
is called a \emph{proper filter}.  For instance, the filter {\tt
  Rbar\_locally p\_infty} contains all intervals of the form \((a,
  +\infty)\) and their supersets, the filter {\tt locally x} contains
all open balls centered in {\tt x} and their supersets, and the
filter {\tt at\_right x} contains the intersections of all members
of {\tt locally x} with the interval \(({\tt x}, +\infty)\).

With two filters \(F_1\) and \(F_2\) on types \(T_1\) and \(T_2\), it
is possible to construct a product filter on \(T_1 \times T_2\), which
contains all cartesian products of a set in \(F_1\) and a set in
\(F_2\) and their supersets.  This corresponds to pairs of points
which tend simultaneously towards the limits described by \(F_1\) and
\(F_2\).

To define a comparison between filters on the real line, we state that
\(F_1\) is less than \(F_2\) if there exists a middle point \(m\), so
that the product filter \(F_1 \times F_2\) accepts the set of pairs \(v_1,
v_2\) such that \(v_1 < m < v_2\).  In other words, this means that
as \(v_1\) tends to \(F_1\) and \(v_2\) to \(F_2\), it ultimately
holds that \(v_1 < m < v_2\).  In yet other words, if there exists an
\(m\) such that the filter \(F_1\) contains \((-\infty, m)\) and
\(F_2\) contains \((m, +\infty)\), then \(F_1\) is less than \(F_2\).
These are expressed by the following definition and the following
theorem:
\begin{verbatim}
Definition filter_Rlt F1 F2 :=
  exists m, filter_prod F1 F2 (fun p => fst p < m < snd p).

Lemma filter_Rlt_witness m (F1 F2  : (R -> Prop) -> Prop) :
  F1 (Rgt m) -> F2 (Rlt m) ->  filter_Rlt F1 F2.
\end{verbatim}
We proved a few comparisons between filters, for instance {\tt at\_right
  \(x\)} is smaller than {\tt Rbar\_locally p\_infty} for any real
\(x\), {\tt at\_left \(a\)} is smaller than {\tt at\_right \(b\)} if \(a
\leq b\), but {\tt at\_right \(c\)} is only smaller than {\tt at\_left \(d\)}
when \(c < d\).

We can reproduce for improper integrals
the results given by the Chasles relations for proper Riemann
integrals.  Here is an example of a Chasles relation:  {\emph if
\(f\) is integrable between \(a\) and \(c\) and \(a \leq b \leq c\),
then
\(f\) is integrable between \(a\) and \(b\) and between \(b\) \and
\(c\), and the integrals satisfy the following relation:}
\[ \int_a^c f(x) \,{\rm d} x = \int_a^b f(x) \,{\rm d} x + \int_b^c f(x)
   \,{\rm d} x\]
This theorem is provided in the Coquelicot library for \(a\), \(b\),
and \(c\) taken as real numbers.  With the order of filters, we
can simply re-formulate this theorem for \(a\) and \(c\) being
arbitrary filters, and \(b\) being a real number between them.  This
is expressed as follows:
\begin{verbatim}
Lemma ex_RInt_gen_cut (a : R) (F G: (R -> Prop) -> Prop)
   {FF : ProperFilter F} {FG : ProperFilter G} (f : R -> R) :
   filter_Rlt F (at_point a) -> filter_Rlt (at_point a) G ->
   ex_RInt_gen f F G -> ex_RInt_gen f (at_point a) G.
\end{verbatim}
We are still considering whether this theorem should be improved,
using the filter {\tt locally \(a\)} instead of {\tt at\_point \(a\)}
for the intermediate integration bound.

The theorem \verb+ex_RInt_gen_cut+ is used three times, once to
establish equation~(\ref{eqn:elliptic-agm-step})  and twice to establish 
equation~(\ref{eqn:elliptic-sqrt-bound}) at page \pageref{eqn:elliptic-agm-step}.

\paragraph{From improper to proper integrals.}  Through variable changes, 
improper integrals can be transformed into proper integrals and
vice-versa.  For instance, the change of variable leading to
equation~(\ref{eqn:elliptic-agm-step}) actually leads to the
correspondence.
\[ \int_0^{+\infty} \frac{{\rm d}t}{\sqrt{(a ^ 2 + t ^ 2)(b ^ 2 + t ^
    2)}} = \frac{1}{2}
\int_{-\infty}^{+\infty} \frac{{\rm d} s}{\sqrt{{((\frac{a +
          b}{2}) ^ 2 + s ^ 2)(ab + s ^ 2)}}} \]
The lower bounds of the two integrals correspond to each other with
respect to the variable change \(s = \frac{1}{2}(t - \frac{ab}{t})\),
but the first lower bound needs to be considered proper for later
uses, while the lower bound for the second integral is necessarily
improper.  To make it possible to change from one to the other, we
establish a theorem that makes it possible to transform a limit bound
into a real one.
\begin{verbatim}
Lemma is_RInt_gen_at_right_at_point (f : R -> R) (a : R) F
  {FF : ProperFilter F v} :
  locally a (continuous f) -> is_RInt_gen f (at_right a) F v ->
  is_RInt_gen f (at_point a) F v.
\end{verbatim}
This theorem contains an hypothesis stating that \(f\) should be well
behaved around the real point being considered, the lower
bound.  In this case, we use an hypothesis of continuity {\em around}
this point, but this hypothesis could probably be made weaker.

\paragraph{Limit equivalence.}  Equations~(\ref{eqn:Iequiv}.1)
and~(\ref{eqn:Mequiv}.2) at page \pageref{eqn:Iequiv} rely on the concept of equivalent functions at a
limit.  For our development, we have not developed a separate concept
for this, instead we expressed statements as the ratio between the
equivalent functions having limit 1 when the input tends to the limit of
interest.  For instance equation~(\ref{eqn:Mequiv}.1) is expressed
formally using the following lemma:
\begin{verbatim}
Lemma M1x_at_0 : filterlim (fun x => M 1 x / (- PI / (2 * ln x)))
                  (at_right 0) (locally 1).
\end{verbatim}
In this theorem, the fact that \(x\) tends to \(0\) on the right is
expressed by using the filter \verb+(at_right 0)+.

We did not develop a general library of equivalence, but we still
gave ourself a tool following the transitivity of this equivalence
relation.  This theorem is expressed in the following manner:
\begin{verbatim}
Lemma equiv_trans F {FF : Filter F} (f g h : R -> R) :
  F (fun x => g x <> 0) -> F (fun x => h x <> 0) ->
  filterlim (fun x => f x / g x) F (locally 1) ->
  filterlim (fun x => g x / h x) F (locally 1) ->
  filterlim (fun x => f x / h x) F (locally 1).
\end{verbatim}
The hypotheses like \verb+F (fun x => g x <> 0)+ express that in
the vicinity of the limit denoted by {\tt F}, the
function should
be non-zero.  The rest of the theorem expresses that if \(f\) is
equivalent to \(g\) and \(g\) is equivalent to \(h\), then \(f\) is
equivalent to \(h\).  To perform this proof, we need to leave the
realm of filters and fall back on the traditional
\(\varepsilon-\delta\) framework.
\paragraph{Uniform convergence and derivatives.}  During our experiments, we
found that the concept of uniform convergence does not fit well in the
framework of filters as provided by the Coquelicot library.  The sensible
approach would be to consider a notion of balls on the space of functions,
where a function \(g\) is inside the ball centered in \(f\) if the
value of \(g(x)\) is never further from the value of \(f(x)\) than the ball
radius, for every \(x\) in the input type.  One would then need to
instantiate the general structures of topology provided by Coquelicot
to this notion of ball, in particular the structures of {\tt
  UniformSpace} and {\tt NormedModule}.  In practice, this does not
provide all the tools we need, because we actually want to restrict the
concept of uniform convergence to subsets of the whole type.  In this
case the structure of {\tt UniformSpace} is still appropriate, but the
concept of {\tt NormedModule} is not, because two functions that
differ outside the considered subset may have distance 0 when
only considering their values inside the subset.

The alternative is provided by a treatment of uniform convergence that
was developed in Coq's standard library of real numbers at the end of
the 1990's, with a notion denoted {\tt CVU \(f\) \(g\) \(c\) \(r\)},
where \(f\) is a sequence of functions from \(\mathbb{R}\) to
\(\mathbb{R}\), \(g\) is a function from \(\mathbb{R}\) to
\(\mathbb{R}\), \(c\) is a number in \(\mathbb{R}\) and \(r\) is
a positive real number.  The meaning is that the sequence of function
\(f\) converges uniformly towards \(g\) inside the ball centered in
\(c\) of radius \(r\).  We needed a formal description of a theorem
stating that when the derivatives \(f'_n\) of a convergent sequence of
functions \(f_n\) tend uniformly to a limit function \(g'\), this function
\(g'\) is the derivative of the limit of the sequence \(f_n\).

There is already a similar theorem in Coq's standard library,
with the following statement:
\begin{verbatim}
derivable_pt_lim_CVU :
forall fn fn' f g x c r,
Boule c r x ->
(forall y n, Boule c r y ->
                          derivable_pt_lim (fn n) y (fn' n y)) ->
(forall y, Boule c r y -> Un_cv (fun n : nat => fn n y) (f y)) ->
CVU fn' g c r ->
(forall y : R, Boule c r y -> continuity_pt g y) ->
derivable_pt_lim f x (g x)
\end{verbatim}
However, this theorem is sometimes impractical to use, because it
requires that we already know the limit derivative to be continuous, a
condition that can actually be removed.  For this reason, we
developed a new formal proof for the theorem, with the following
statement\footnote{It turns out that the theorem
{\tt derivable\_pt\_lim\_CVU} was already introduced by a previous
study on the implementation of \(\pi\) in the Coq standard library
of real numbers \cite{BertotAllais14}.}
\begin{verbatim}
Lemma CVU_derivable :
forall f f' g g' c r,
 CVU f' g' c r ->
 (forall x, Boule c r x -> Un_cv (fun n => f n x) (g x)) ->
 (forall n x, Boule c r x ->
                      derivable_pt_lim (f n) x (f' n x)) ->
 forall x, Boule c r x -> derivable_pt_lim g x (g' x).
\end{verbatim}
In this theorem's statement, the third line expresses that the
derivatives {\tt f'} converge uniformly towards the function {\tt g'},
the fourth line expresses that the functions {\tt f} converge simply
towards the function {\tt g} inside the ball of center {\tt c} and
radius {\tt r}, the fifth and sixth line express
  that the functions {\tt f}
are differentiable everywhere inside the ball and the derivative is {\tt
  f'}, and the seventh line concludes that the function {\tt g} is
differentiable everywhere inside the ball and the derivative is {\tt g'}.
While most of the theorems we wrote are expressed using concepts from
the Coquelicot library, this one is only expressed with concepts
coming from Coq's standard library of real numbers, but all these
concepts, apart from {\tt CVU}, have a Coquelicot equivalent (and
Coquelicot provides the foreign function interface): {\tt Boule c r x}
is equivalent to {\tt Ball c r x} in Coquelicot, {\tt Un\_cv \(f\) \(l\)} is
equivalent to {\tt filterlim \(f\) Eventually (locally \(l\))}, and
\verb+derivable_pt_lim+ is equivalent to \verb+is_derive+.

We used the theorem \verb+CVU_derivable+ twice in our development,
once to
establish that function \(x \mapsto M(1, x)\) is differentiable everywhere
in the open interval \((0,1)\) and the sequence of derivatives of the
\(a_n\) functions converges to its derivative, and once to
establish that the derivatives of the \(k_n\) functions converge
to \(M^2(1, x) / (x (1 - x ^ 2))\), as in
equation~(\ref{eqn:main_derivative}).

\paragraph{Automatic proofs.}  In this development, we make an extensive
use of divisions and square root.  To reason about these functions, it
is often necessary to show that the argument is non-zero (for division),
or positive (for square root).  There are very few automatic tools
to establish this kind of results in general about real numbers, especially
in our case, where we rely on a few transcendental functions.  For
linear arithmetic formulas, there exists a tool call {\tt psatzl R}
\cite{DBLP:conf/types/Besson06},
that is very useful and robust in handling of conjunctions and its use of
facts from the current context.  Unfortunately, we have many expressions
that are not linear.  We decided to implement a semi-automatic tactic
for the specific purpose of proving that numbers are positive, with the
following ordered heuristics:
\begin{itemize}
\item Any positive number is non-zero,
\item all exponentials are positive,
\item \(\pi\), \(1\), and \(2\) are positive,
\item the power, inverse, square root of positive numbers is positive,
\item the product of positive numbers is positive,
\item the sum of an absolute value or a square and a positive number is positive,
\item the sum of two positive numbers are positive,
\item the minimum of two positive numbers is positive,
\item a number recognized by the {\tt psatzl R} tactic to be positive
is positive.
\end{itemize}
This semi-automatic tactic can easily be implemented using Coq's tactic
programming language {\tt Ltac}.  We named this tactic {\tt lt0} and it
is used extensively in our development.

Given a function
like \(x \mapsto 1 / \sqrt{(x ^ 2 + a ^ 2)(x ^ 2 + b ^ 2)}\), the
Coquelicot library provides automatic tools (mainly a tactic called {\tt
  auto\_derive}) to show that this function
is differentiable under conditions that are explicitly computed.  For this
to work, the tool needs
to rely on a database of facts concerning all functions involved.  In
this case, the database must of course contain facts about
exponentiation, square roots, and the inverse function.  As a result,
the tactic {\tt auto\_derive} produces conditions, expressing
that \((x ^ 2 + a ^ 2)(x ^ 2 + b ^ 2)\) must be positive and the
whole square root expression must be non zero.

The tactic {\tt auto\_derive} is used more than 40 times in our
development, mostly because there is no automatic tool to show
the continuity of functions and we rely on a theorem that states
that any differentiable function is continuous, so that we often prove derivability
only to prove continuity.

When proving that the functions \(a_n\) and \(b_n\) are
differentiable, we need to rely on a more elementary proof tool, called
{\tt auto\_derive\_fun}.  When given a function to derive, which
contains functions that are not known in the database, it builds an
extra hypothesis, which says that the whole expression is differentiable as
soon as the unknown functions are differentiable.  This is especially
useful in this case, because the proof that \(b_n\) is differentiable
is done recursively, so that there is no pre-existing theorem stating
that \(a_n\) and \(b_n\) are differentiable when studying the derivative of
\(b_{n+1}\).  For instance, we can call the following tactic:
\begin{verbatim}
auto_derive_fun (fun y => sqrt (a_ n y * b_ n y)); intros D.
\end{verbatim}
This creates a new hypothesis named {\tt D} with the following
statement:
\begin{verbatim}
 D : forall x : R,
    ex_derive (fun x0 : R => a_ n x0) x /\
    ex_derive (fun x0 : R => b_ n x0) x /\
    0 < a_ n x * b_ n x /\ True ->
    is_derive (fun x0 : R => sqrt (a_ n x0 * b_ n x0)) x
      ((1 * Derive (fun x0 : R => a_ n x0) x * b_ n x +
        a_ n x * (1 * Derive (fun x0 : R => b_ n x0) x)) *
       / (2 * sqrt (a_ n x * b_ n x)))
\end{verbatim}

Another place where automation provides valuable help is when we wish
to find good approximations or bounds for values.  The {\tt interval}
tactic \cite{interval} works on goals consisting of such comparisons
and solves them
right away, as long as it knows about all the functions involved.
Here is an example of a comparison that is easily solved by this
tactic:
\begin{verbatim}
   (1 + ((1 + sqrt 2)/(2 * sqrt (sqrt 2))))
          / (1 + / sqrt (/ sqrt 2)) < 1
\end{verbatim}

An example of expression where {\tt interval} fails, is when the expressions
being considered are far too large.  In our case, we wish to prove
that
\[ 4 \pi_0 \frac{1}{531 ^ {2 ^ {19}}} \leq \frac{1}{10 ^{10 ^ 6 + 4}}\]
The numbers being considered are too close to 0 for {\tt interval} to work.

\label{sec:logarithmic-downscale}
The solution to this problem is to first use monotonicity properties
of either the logarithm function (in the current version of our
development) or the exponential function (in the first version), thus
resorting to symbolic computation before finishing off with the
{\tt interval} tactic.

The {\tt interval} tactic already knows about the \(\pi\) constant, so
that it is
quite artificial to combine our result from formula~(\ref{eqn:bound_pi})
and this
tactic to obtain approximations of \(\pi\) but we can still make this
experiment and establish that the member \(\pi_3\) of the sequence
is a good enough approximation to know all first 10 digits of \(\pi\).
Here is the statement:
\begin{verbatim}
Lemma first_computation :
   3141592653/10 ^ 9 < agmpi 3 /\
     agmpi 3 + 4 * agmpi 0 * Rpower 531 (- 2 ^ 2)
    < 3141592654/10 ^ 9.
\end{verbatim}
We simply expand fully {\tt agmpi}, simplify instances of \(y_n\)
and \(z_n\) using the 
equations~(\ref{eqn:y_step_z_step}), and then ask the {\tt interval}
tactic to finish the comparisons.  We need to instruct
the tactic to use 40 bits of precision.  This takes some time (about a
second for each of the two comparisons), and we conjecture that the expansion of
all functions leads to sub-expression duplication, leading also to
duplication of work.  When aiming for more distant decimals, we will need
to apply another solution.
\section{Computing large numbers of decimals}
Theorem provers based on type theory have the advantage that they provide
computation capabilities on inductive types.  For instance, the Coq system
provides
a type of integers that supports comfortable computations for integers
with size going up to \(10 ^{100}\).  Here is an example computation, which
feels instantaneous to the user.
\begin{verbatim}
Compute (2 ^331)%Z.
     = 174980057982640953949800178169409709228253554471456994914
  06164851279623993595007385788105416184430592
     : Z
\end{verbatim}
By their very nature, real numbers cannot be provided as an inductive datatype
in type theory.  Thus the {\tt Compute} command will not perform any
computation for the similar expression concerning real numbers.  The reason is
that while some real numbers are defined like integers by applying simple
finite operations on basic constants like 0 and 1, other are only obtained
by applying a limiting process, which cannot be represented by a finite
computation.  Thus, it does not make sense to ask to compute an expression
like \(\sqrt{2}\) in the real numbers, because there is no way to provide
a better representation of this number than its definition.  On the other
hand, what we usually mean by \emph{computing \(\sqrt{2}\)} is to provide
a suitable approximation of this number.  This is supported in the Coq system
by the {\tt interval} tactic, but only when we are in the process of
constructing
a proof, as in the following example:
\begin{verbatim}
Lemma anything : 12 / 10 < sqrt 2.
Proof.
interval_intro (sqrt 2).


1 subgoal

  H : 759250124 * / 536870912 <= sqrt 2 <= 759250125 * / 536870912
  ============================
   12 / 10 < sqrt 2
\end{verbatim}
What we see in this dialog is that the system creates a new hypothesis
(named {\tt H}) that provides a new fact giving an approximation of
\(\sqrt{2}\).  In this hypothesis, the common numerator appearing
in both fractions is actually the number \(2 ^{29}\).  Concerning
notations, readers will have to know that Coq writes \hbox{\tt/ a} for
the inverse of {\tt a}, so that \hbox{\tt 3 * / 2} is 3 times the inverse
of 2.  A human mathematician would normally write \hbox{\tt 3 / 2}
and Coq also accepts this syntax.

One may argue that {\tt 759250124 * / 536870912} is not much better
than {\tt sqrt 2} to represent that number, and actually this ratio
is not exact,
but it can be used to help proving that \(\sqrt{2}\) is larger or
smaller than another number.

Direct computation on the integer datatype can also be used to
approximate computations in real numbers.  For instance, we can
compute the same numerator for an approximation of \(\sqrt{2}\) by
computing the integer square root of \(2 \times (2 ^ {29}) ^2\).

\begin{verbatim}
Compute (Z.sqrt (2 * (2 ^ 29) ^ 2)).
     = 759250124%Z
     : Z
\end{verbatim}
This approach of computing integer values for numerators of rational numbers
with a fixed denominator is the one we are going to
exploit to compute the first million digits of \(\pi\), using three
advantages provided by the Coq system:
\begin{enumerate}
\item The Coq system provides an implementation of {\em big integers}, which
can withstand computations of the order of \(10 ^{10 ^{12}}\).
\item The big integers library already contains an efficient implementation of integer square roots.
\item The Coq system provides a computation methodology where code is compiled
into OCaml and then into binary format for fast computation.
\end{enumerate}
\subsection{A framework for high-precision computation}
\label{rounding-big}
If we choose to represent every computation on real numbers by a
computation on corresponding approximations of these numbers, we need
to express how each operation will be performed and interpreted.  We
simply provide five values and functions that implement the elementary
values of \(\mathbb{R}\) and the elementary operations:
multiplication, addition, division, the number 1, and the number 2.

We choose to represent the real number \(x\) by the integer \(\lfloor
m x\rfloor\) where \(m\) is a scaling factor that is mostly fixed for
the whole computation.  For readability, it is often practical to use
a power of 10 as a scaling factor, but in this paper, we will also see
that we can benefit from also using scaling factors that are powers of
2 or powers of 16.  Actually, it is not even necessary that the
scaling factor be any power of a small number, but it turns out that
it is the most practical case.

Conversely, we shall note \(\denote{n}\) the real value represented by
the integer \(n\).  Simply, this number is \(\frac{n}{m}\).

When \(m\) is the scaling factor, the real number \(1\) is represented by
the integer \(m\) and the real number \(2\) is represented by the number
\(2\times m\).  So \(\denote{m} = 1\), \(\denote{2m} = 2\).  So,
we define the following two functions to describe the representations
of \(1\) and \(2\) with respect to a given scaling factor, in Coq
syntax where we use the name {\tt magnifier} for the scaling factor.
\begin{verbatim}
Definition h1 (magnifier : bigZ) := magnifier.
Definition h2 magnifier := (2 * magnifier)%bigZ.
\end{verbatim}

When multiplying two real numbers \(x\) and \(y\), we
need to multiply their representations and take care of the scaling.
To understand how to handle the scaling, we should look at the
following equality:
\[\denote{n_1} \denote{n_2} = \frac{n_1}{m}\frac{n_2}{m}\]
To obtain the integer that will represent this result, we need to
multiply the product of the represented numbers  by \(m\) and then take the largest integer below.  This
is
\[\lfloor \frac{n_1 \times n_2}{m}\rfloor\]
The combination of the division operation and taking the largest
integer below is performed by integer division.  So we define our
high-precision multiplication as follow.
\begin{verbatim}
Definition hmult (magnifier x y : bigZ) :=
  (x * y / magnifier)%bigZ.
\end{verbatim}
For division and square root, we reason similarly.

For addition, nothing needs to be implemented, we can directly use
integer computation.  The scaling factor is transmitted naturally (and
linearly from the operands to the result).  Similarly, multiplication
by an integer can be represented directly with integer multiplication,
without having to first scale the integer.

Here are a few examples.  To compute \(\frac{1}{3}\) to a precision
of \(10 ^{-5}\), we can run the following computation.
\begin{verbatim}
Compute let magnifier := (10 ^ 5)%bigZ in
  hdiv magnifier magnifier (3 * magnifier).
     = 33333%bigZ
     : BigZ.t_
\end{verbatim}
The following illustrates how to compute \(\sqrt{2}\) to the same precision.
\begin{verbatim}
Compute let magnifier := (10 ^ 5)%bigZ in
  hsqrt magnifier (2 * magnifier).
     = 141421%bigZ
     : BigZ.t_
\end{verbatim}
In both examples, the real number of interest has the order of magnitude
of \(1\) and is represented by a 5 or 6 digit integer.  When we want
to compute one million decimals of \(\pi\) we should handle integers
whose decimal representation has approximately one million digits.
Computation with this kind of numbers takes time.  As an example, we propose
a computation that handles the 1 million digit representation of
\(\sqrt{2}\) and avoids displaying this number (it only checks that
the millionth decimal is odd).
\begin{verbatim}
Time Eval  native_compute in
   BigZ.odd (BigZ.sqrt (2 * 10 ^ (2 * 10 ^ 6))).
     = true
     : bool
Finished transaction in 91.278 secs (90.218u,0.617s) (successful)
\end{verbatim}
This example also illustrates the use of a different evaluation strategy
in the Coq system, called {\tt native\_compute}.  This evaluation
strategy relies on compiling the executed code in OCaml and then on
relying on the most efficient variant of the OCaml compiler to produce
a code that is executed and whose results are integrated in the memory
 of the Coq system \cite{full-throttle}.  This strategy relies on the
OCaml compiler and the operating system linker in ways that are more
demanding than traditional uses of Coq.  Still it is the same compiler
that is being used as the one used to compile the Coq system, so that the
trusted base is not changed drastically in this new approach.

When it comes to time constraints, all scaling factors are not
as efficient.  In conventional computer arithmetics, it is well-known that
multiplications by powers of 2 are less costly, because they can
simply be implemented by shifts on the binary representation of
numbers.  This property is also true for Coq's implementation of big
integers.  If we compare
the computation of \(\sqrt{\sqrt{2}}\) when the scaling factor is
\(10 ^{10 ^ 6}\) or \(2 ^ {3321929}\), we get a
performance ratio of 1.5, the latter setting is faster even though
the scaling factor and the intermediate values are slightly larger.

It is also interesting to understand how to stage computations, so
that we avoid performing the same computation twice.  For this
problem, we have to be careful, because values that are precomputed
don't have the same size as their original description, and this
may not be supported by the {\tt native\_compute} chain of evaluation.
Indeed, the following experiment fails.
\begin{verbatim}
Require Import BigZ.

Definition mag := Eval native_compute in (10 ^ (10 ^ 6))%bigZ.

Time Definition z1 := Eval native_compute in
   let v := mag in (BigZ.sqrt (v * BigZ.sqrt (v * v * 2)))%bigZ.
\end{verbatim}
This examples makes Coq fail, because the definition of {\tt mag} with
the pragma {\tt Eval native\_compute in} makes that the value
\(10 ^ {10 ^ 6}\) is precomputed, thus creating a huge object
of the Gallina language, which is then passed as a program for the
OCaml compiler to compile when constructing {\tt z1}.  The compiler
fails because the input program is too large.

On the other hand, the following computation succeeds:
\begin{verbatim}
Eval native_compute in
   let v := (10 ^ (10 ^ 6))%bigZ in
   (BigZ.sqrt (v * BigZ.sqrt (v * v * 2))).
\end{verbatim}

\subsection{The full approximating algorithm}
Using all elementary operations described in the previous section,
we can describe the recursive algorithm to compute approximations
of \(\pi_n\) in the following manner.
\begin{verbatim}
Fixpoint hpi_rec (magnifier : bigZ)
  (n : nat) (s2 y z prod : bigZ) {struct n} : bigZ :=
  match n with
  | 0%nat =>
      hmult magnifier (h2 magnifier + s2) prod
  | S p =>
      let sy := hsqrt magnifier y in
      let ny := hdiv magnifier (h1 magnifier + y) (2 * sy) in
      let nz :=
        hdiv magnifier (h1 magnifier + hmult magnifier z y)
          (hmult magnifier (h1 magnifier + z) sy) in
      hpi_rec magnifier p s2 ny nz
        (hmult magnifier prod
           (hdiv magnifier (h1 magnifier + ny)
                           (h1 magnifier + nz)))
  end.
\end{verbatim}
This function takes as input the scaling factor {\tt magnifier},
 a number of iteration {\tt n},
the integer {\tt s2} representing \(\sqrt{2}\), the integer {\tt y}
representing
\(y_p\) for some natural number \(p\) larger than 0, the integer
{\tt z} representing \(z_p\), and the integer {\tt prod} representing
the value
\[\prod_{i=1}^{p} \frac{1 + y_i(\frac{1}{\sqrt{2}})}{1 +
  z_i(\frac{1}{\sqrt{2}})}\]
It computes an integer approximating \(\pi_{n+p} \times {\tt magnifier}\),
but not exactly this number.  The number {\tt s2} is passed
as an argument to make sure it is not computed twice, because it is
already needed to compute the initial values of {\tt y}, {\tt z},
 and {\tt prod}.  This
recursive function is wrapped in the following functions.
\begin{verbatim}
Definition hs2 (magnifier : bigZ) :=
  hsqrt magnifier (h2 magnifier).

Definition hsyz (magnifier : bigZ) :=
  let hs2 := hs2 magnifier in
  let hss2 := hsqrt magnifier hs2 in
  (hs2, (hdiv magnifier (h1 magnifier + hs2) (2 * hss2)), hss2).

Definition hpi (magnifier : bigZ) (n : nat) :=
match n with
| 0%nat =>
    (h2 magnifier + (hs2 magnifier))%bigZ
| S p =>
    let '(s2, y1 , z1) := hsyz magnifier in
    hpi_rec magnifier p s2 y1 z1
      (hdiv magnifier (h1 magnifier + y1)
         (h1 magnifier + z1))
end.
\end{verbatim}
We can use this function {\tt hpi} to compute approximations of \(\pi\) at a
variety of precisions.  Here is a collection of trials performed on
a powerful machine.
\begin{center}
\begin{tabular}{|l|c|c|c|c|}
\hline
scale(iterations)& \(10 ^ {10 ^ 4} (14)\) & \(2 ^ {33220} (14)\) & \(10 ^ {10 ^ 5}(17)\) &
\(2 ^{332193}(17)\) \\
\hline
time & 9s & 4s & 5m30s & 2m30s\\
\hline
\end{tabular}
\end{center}
This table illustrates the advantage there is to compute with a
scaling factor that is a power of 2.  Each column where the scaling
factor is a power of 2 gives an approximation that is slightly more
precise than the column to its left, at a fraction of the cost in
time.  Even if our objective is to
obtain {\em decimals} of \(\pi\), it should be efficient to first perform
the computations of all the iterations with a magnifier that is a
power of 2, only to change the scaling factor at the end of the
computation, this is the solution we choose eventually.

There remains a question about how much precision is lost when so many
computations are performed with elementary operations that each
provide only approximations of the mathematical operation.
Experimental evidence shows that when computing 17 iterations with a
magnifier of \(10 ^ {10 ^ 5}\) the last two digits are wrong.  The
next section shows how we prove bounds on the accumulated error in the
concrete computation.
\section{Proofs about approximate computations}
When proving facts about approximate computations, we want to abstract
away from the fact that the computations are performed with a
datatype that provides fast computation with big integers.  What
really matters is that we approximate each operation on real numbers
by another operation on real numbers and we have a clear description
of how the approximation works.  In the next section, we describe
the abstract setting and the proofs performed in this setting.  In
a later section, we show how this abstract setting is related to the
concrete setting of computing with integers and with the particular
datatype of big integers.

\subsection{Abstract reasoning on approximate computations}
\label{sec:abstract-approximate}
In the case of fixed precision computation as we described in the
previous section, we know that all operations are approximated from
below by a value which is no further than a fixed allowance \(e\).
This does not guarantee that all values are approximated from below,
because one of the approximated operations is division, and dividing
by an approximation from below may yield an approximation from above.

For this reason, most of our formal proofs about approximations are
performed in a section where we assume the existence of a collection
of functions and their properties.

The header of our working section has the following content.
\begin{verbatim}
Variables (e : R) (r_div : R -> R -> R) (r_sqrt : R -> R)
           (r_mult : R -> R -> R).

Hypothesis ce : 0 < e < /1000.

Hypothesis r_mult_spec :
  forall x y, 0 <= x -> 0 <= y ->
   x * y - e < r_mult x y <= x * y.
\end{verbatim}
In this header, we introduce a constant {\tt e}, which is used to bound
the error made in each elementary operation, we assume that {\tt e}
is positive and suitably small, and then we describe how each
rounded operation behaves with respect to the mathematical operation
it is supposed to represent.  For multiplication, the hypothesis named
{\tt r\_mult\_spec} describes that the inputs are expected to be
positive numbers, and that the result of {\tt r\_mul x y} is smaller
than or equal to the product, but the difference is smaller than {\tt
  e} in absolute value.  We have similar specification hypotheses
for the rounded division {\tt r\_div} and the rounded square root
{\tt r\_sqrt}.  We then use these rounded operations to describe the
computations performed in the algorithm.

We can now study how the computation of the various sequences of
the algorithm are rounded, and how errors accumulate.  Considering
the sequence \(y_n\), the computation at each step is represented
by the following expression.
\begin{verbatim}
r_div (1 + y) (2 * (r_sqrt y))
\end{verbatim}
In this expression, we have to assume that {\tt y} comes from a
previous computation, and for this reason it is tainted with some
error {\tt h}.  The question we wish to address has the following
form: {\em if we know that \(y_n\) is tainted with an error \(h\)
that  is smaller that a given allowance \(e'\), can we show
that \(y_{n+1}\) is tainted with an error that is smaller than
\(f(e')\) for some well-behaved function \(f\)?  How much bigger than
\(e\) must \(e'\) be?}

We were able to answer two questions:
\begin{itemize}
\item if the accumulated error on computing \(y_n\) is smaller than \(e'\),
  then the accumulated error on computing \(y_{n+1}\) is also smaller than
  \(e'\) (so for the sequence \(y_n\), the function \(f\) is the
  identity function),
\item the allowance \(e'\) needs to be at least \(2 e\) (and not more).
\end{itemize}
This is quite surprising. Errors don't really accumulate for
this sequence.

In retrospect, there are good reasons for this.  Rounding
errors in the division operation make the result go down, but
rounding errors in the square root make the result go up.
On the other hand, the input value \(y_n\) may be tainted by an error
\(h\), but this error is only multiplied by the derivative of the function
\[y \mapsto \frac{1 + y}{2 \sqrt{y}}\]
It happens that this derivative never exceeds \(\frac{1}{14}\) in the
region of interest.

As an illustration, let's assume \(y_n = 1.100\), we want to compute
\(y_{n+1}\), and we are working with three digits of precision.  The value
of \(\sqrt{1.1}\) is \(1.04880\dots\) but it is rounded down to \(1.048\).
\(2\sqrt{1.1}\) is \(1.09761\dots\) but the rounded computation give \(2.096\),
\(y_{n+1}\) is \(1.00113\).  In our computation, we actually compute \((1 + 1.1)/2.096)=1.00190\).  This is an over approximation of \(y_{n+1}\), but this is
rounded down to \(1.001\): the last rounding down compensates the
over-approximation introduced when dividing by the previously rounded
down square root.  If our input representation of \(y_n\) is an
approximation, for example we compute with \(1.098\) or \(1.102\), we still
obtain \(1.001\).

In the end, the lemma we are able to prove has the following
statement.
\begin{verbatim}
Lemma y_error e' y h :
  e' < /10 -> e <= e' / 2 -> 1 <= y <= 71/50 -> Rabs h < e' ->
  let y1 := (1 + y)/(2 * sqrt y) in
  y1 - e' < r_div (1 + (y + h)) (2 * (r_sqrt (y + h))) < y1 + e'.
\end{verbatim}
The proof is organized in four parts, where the first part consists
in replacing the operations with rounding by expressions where
an explicit error ratio is displayed.  We basically construct 
a value {\tt e1}, taken in the interval \([-\frac{1}{2},0]\),
so that the following equality holds.
\begin{verbatim}
r_sqrt (y + h) = sqrt (y + h) + e1 *  e'
\end{verbatim}
We prefer to define {\tt e1} as a ratio between constant bounds,
rather than a value in an interval whose bounds are expressed in {\tt
  e'}, because the automatic tactic {\tt interval} handles values
between numeric constants better.  We do the same for the division, introducing
a ratio {\tt e2}.

The second part of the proof consists
in showing that the propagated error from previous computations has
limited impact on the final error.  This is stated as follows.
\begin{verbatim}
set (y2 := (1 + (y + h)) / (2 * sqrt (y + h))).
assert (propagated_error : Rabs (y2 - y1) < e' / 14).
\end{verbatim}
This step is proved by applying the mean value theorem, using the
derivative of the function \(y \mapsto \frac{1 + y}{2\sqrt{y}}\),
which was already computed during the proof of convergence of the
\(y_n\) sequence.  The {\tt interval} tactic is practical here
to show the absolute value of the derivative of that
function at any point
between {\tt y} and {\tt y + h} is below \(\frac{1}{14}\).  The mean
value theorem makes it possible to factor out the input error in the
comparisons, so that we eventually obtain a comparison of an
expression with a constant, which we resolve using the {\tt interval}
tactic.

The other two parts of the proof are concerned with providing a bound
for the impact of the rounding errors introduced by the current
computation.  Each part is concerned with one direction, and in each
case only one of the two possible rounding errors need to be
considered.

The proof for the lemma {\tt y\_error} is quite long (just under 100
lines), but this is only a preliminary step for the proof of lemma
{\tt z\_error}, which shows that the errors accumulated when computing
the \(z_n\) sequence can also be bounded in a constant fashion.  The
statement of this lemma has the following shape.
\begin{verbatim}
Lemma z_error e' y z h h' :
  e' < /50 -> e <= e' / 4 -> 1 < y < 51/50 -> 1 < z < 6/5 ->
  Rabs h < e' -> Rabs h' < e' ->
  let v := (1 + z * y)/((1 + z) * sqrt y) in
  v - e' < r_div (1 + r_mult (z + h') (y + h))
            (r_mult (1 + (z + h')) (r_sqrt (y + h))) < v + e'.
\end{verbatim}
In this statement, the fragment
\begin{verbatim}
   r_div (1 + r_mult (z + h') (y + h))
       (r_mult (1 + (z + h') (r_sqrt (y + h)))
\end{verbatim}
represents the computed expression with rounding operations, using
inputs that are tainted by errors {\tt h} and {\tt h'}, while
the fragment
\begin{verbatim}
(1 + z * y) /((1 + z) * sqrt y)
\end{verbatim}
represents the ratio \(\frac{1 + zy}{(1 + z)\sqrt{y}}\).

This proof is more complex.  In this case, we are also able to
show that errors do not grow as we compute more elements of
the sequence: they stay stable at about 4 times the elementary
rounding error introduced by each rounding operation.  The
proof of this lemma is around 170 lines long.

The next step in the computation is to compute the product of
ratios \(\prod \frac{1 + y}{1 + z}\).  For each ratio, we
establish a bound on the error as expressed by the following
lemma.
\begin{verbatim}
Lemma quotient_error : forall e' y z h h', e' < / 40 ->
  Rabs h < e' / 2 -> Rabs h' < e' -> e <= e' / 4 ->
  1 < y < 51 / 50 -> 1 < z < 6 / 5 ->
  Rabs (r_div (1 + (y + h)) (1 + (z + h')) -
              (1 + y)/(1 + z)) <  13 / 10 * e'.
\end{verbatim}
The difference between the second hypothesis (on {\tt Rabs h})
and the third hypothesis {\tt Rabs h'} handles the fact that
we don't have as precise a bound on error for the computation of
\(y_n\) and for \(z_n\).  The result is that the error on the
ratio is bounded at a value just above 5 times the elementary error
{\tt e}.

It remains to prove a bound on the error introduced when computing
the iterated product.  This is done by induction on the number of
iterations.
The
following lemma is used as the induction step: when {\tt p} represents
the product of \(k\) terms and {\tt v} represents one of the
ratios, the product of {\tt p} and {\tt v} with accumulated errors,
adding the error for the rounded multiplication increases by
\(\frac{23}{20}\) the error on the ratio.
\begin{verbatim}
Lemma product_error_step :
  forall p v e1 e2 h h', 0 <= e1 <= /100 -> 0 <= e2 <= /100 ->
    e < /5 * e2 -> /2 < p < 921/1000 ->
    /2 < v <= 1 -> Rabs h < e1 -> Rabs h' < e2 ->
    Rabs (r_mult (p + h) (v + h') - p * v) < e1 + 23/20 * e2.
\end{verbatim}
At this point we write functions {\tt rpi\_rec} and {\tt rpi}
so that they mirror exactly the functions {\tt hpi\_rec} and
{\tt hpi}.  The main difference is that {\tt rpi\_rec} manipulates
real numbers while {\tt hpi\_rec} manipulates integers.  Aside
from this, {\tt rpi\_rec} performs a multiplication using {\tt r\_mult}
wherever {\tt hpi\_rec} performs a multiplication using {\tt hmult}.

We can now combine all results about the sub-expressions, scale
all errors with respect to the elementary error, and obtain a bound
on accumulated errors in {\tt rpi\_rec}, as expressed in the following
lemma.
\begin{verbatim}
Lemma rpi_rec_correct (p n : nat) y z prod :
    (1 <= p)%nat -> 4 * (3/2) * (p + n) * e < /100 ->
    Rabs (y - y_ p (/sqrt 2)) < 2 * e ->
    Rabs (z - z_ p (/sqrt 2)) < 4 * e ->
    Rabs (prod - pr p) < 4 * (3/2) * p * e ->
    Rabs (rpi_rec n y z prod - agmpi (p + n)) <
      (2 + sqrt 2) * 4 * (3/2) * (p + n) * e + 2 * e.
\end{verbatim}
Note that this statement guarantees a bound on errors only if the
magnitude of the error {\tt e} is small enough when compared with
the inverse of the number of iterations {\tt p + n}.  In practice,
this is not a constraint because we tend to make the error magnitude
vanish twice exponentially.

In the end, we have to check the approximations for the initial values
given as argument to {\tt rpi\_rec}.  This
yields a satisfying rounding error lemma.
\begin{verbatim}
Lemma rpi_correct : forall n, (1 <= n)%nat -> 6 * n * e < /100 ->
  Rabs (rpi n - agmpi n) < (21 * n + 2) * e.
\end{verbatim}
In other words, we can
guarantee that \(\pi_n\) is computed with an error that grows
proportionally to \(21 n + 2\).

A similar study for the {\em Brent-Salamin} algorithm yields the following
error estimate:
\begin{verbatim}
Lemma rsalamin_correct (n : nat) :
 0 <= e <= / 10 ^ (n + 6) / 3 ^ (n + 1) ->
  Rabs (rsalamin n - salamin_formula (n + 1)) <=
  (160 * (3 / 2) ^ (n + 1) + 80 * 3 ^ (n + 1) + 100) * e.
\end{verbatim}
This error grows exponentially with respect to \(n\), which means that
the number of needed extra digits to ensure a given distant decimal is
still linear in \(n\).  When computing the number of required extra
digits for 1 million, we obtain 12 (because \(n\) is 19).
\subsection{From abstract rounding to integer computations}
In our concrete setting, we don't have the functions {\tt r\_mult},
{\tt r\_div}, and {\tt r\_sqrt}, but functions {\tt hmult}, {\tt hdiv}
and {\tt hsqrt}.  The type on which these functions operate is {\tt
  bigZ}, a type that is designed to make large computations possible
inside the Coq system, but that is otherwise not suited to perform
intensive proofs.  To establish the connection with our proofs of
rounded operations, we build a bridge that relies on the better
supported type {\tt Z}.

The standard library of reals already provides function {\tt INR} and
{\tt IZR} to inject natural numbers and integers, respectively, into the
type of real numbers.  These functions are useful to us, but they must
be improved to include the scaling process.

We also define functions {\tt hR : Z -> R} and
{\tt Rh : R   -> Z} mapping an integer (respectively a real number)
to its representation (respectively to the integer that represents its
rounding by default).  All these functions are defined in the context
of a Coq section where we assume the existence of a scaling factor
named {\tt magnifier} (an integer), and that this scaling factor is
larger than 1000, which corresponds to assuming that we perform
computations with at least 3 digits of precision.

Coming from the type of integers, we can now redefine the functions {\tt
  hmult}, {\tt hdiv}, and {\tt hsqrt} as in
section~\ref{rounding-big}, but with the type {\tt Z} for inputs and
outputs.  

\begin{verbatim}
Definition hR (v : Z) : R := (IZR v /IZR magnifier)%R.

Definition RbZ (v : R) : Z := floor v.

Definition Rh (v : R) : Z := RbZ( v * IZR magnifier).
\end{verbatim}
The abstract functions {\tt r\_mult}, {\tt r\_div} and {\tt r\_sqrt}
are then defined by rounding and injecting the result back into
the type of real numbers.
\begin{verbatim}
Definition r_mult (x y : R) : R := hR (Rh (x * y)).
\end{verbatim}
The main rounding property can be proved once for all three rounded
operations, since it is solely a property of the {\tt hR} and {\tt Rh}
function.
\begin{verbatim}
Lemma hR_Rh (v : R) : v - /IZR magnifier < hR (Rh v) <= v.
\end{verbatim}

The link to the concrete computing functions is established by the
following kind of lemma, the form of which is close to a morphism
lemma.
\begin{verbatim}
Lemma hmult_spec :
  forall x y : Z, (0 <= x -> 0 <= y ->
   hR (hmult x y) = r_mult (hR x) (hR y))%Z.
\end{verbatim}
The hypotheses {\tt r\_mult\_spec}, {\tt r\_div\_spec}, and
{\tt r\_sqrt\_spec}, which are necessary for the abstract reasoning
in section~\ref{sec:abstract-approximate}, are then easily obtained
by composing a lemma of the form {\tt hmult\_spec} with the lemma
{\tt hR\_Rh}.

The complement of the lemma {\tt hR\_Rh} is another lemma
which expresses that {\tt Rh} is a left
inverse to {\tt hR}.  This lemma is instrumental when showing the
correspondence between concrete and abstract algorithms.

We now have two views of the algorithm: the algorithm {\tt hpi} as
described in section~\ref{rounding-big} and the algorithm {\tt rpi}
where the functions {\tt hmult}, {\tt hdiv}, {\tt hsqrt} have been
replaced by {\tt r\_mult}, {\tt r\_div}, {\tt r\_sqrt} respectively.
We wish to show that these algorithms actually describe the same
computation.  A new difficulty arises because we need to show that all
operations receive and produce non-negative numbers, because these
conditions are required by lemmas like {\tt hmult\_spec}.  This is
not as simple as it seems because the result of {\tt hmult 0 0} is
only guaranteed to be larger than {\tt - e} by the initial specification.
The implementation actually satisfies a stronger property.

In the end the correspondence lemma has the following form.
\begin{verbatim}
Lemma hpi_rpi_rec n p y z prod:
    (1 <= p)%nat ->
    4 * (3/2) * INR (p + n) * /IZR magnifier < /100 ->
    Rabs (hR y - y_ p (/sqrt 2)) < 2 * /IZR magnifier ->
    Rabs (hR z - z_ p (/sqrt 2)) < 4 * /IZR magnifier ->
    Rabs (hR prod - pr p) < 4 * (3/2) * INR p * /IZR magnifier ->
    hR (hpi_rec n y z prod) =
    rpi_rec r_div r_sqrt r_mult n (hR y) (hR z) (hR prod).
\end{verbatim}
The interesting part of this lemma is the equality stated on the last
two lines.  The previous lines only state information about the
size of the inputs, to help make sure that the intermediate
computations never feed a negative number to the operations.  This
constraint of non-negative operands makes the proof of correspondence
tedious, but quite regular.  This proof ends up being 120 lines
long.\footnote{In retrospect, it might have been useful to add hypotheses
that returned values by all functions were positive, as long as the
inputs were.}

A similar proof is constructed for the main encapsulating function, so
that we obtain a lemma of the following shape.
\begin{verbatim}
Lemma hpi_rpi (n : nat) :
  6 * INR n * /IZR magnifier < / 100 ->
  hR (hpi n) = rpi r_div r_sqrt r_mult n.

Lemma integer_pi :
  forall n, (1 <= n)%nat ->
  600 * INR (n + 1) < IZR magnifier < Rpower 531 (2 ^ n)/ 14 ->
  Rabs (hR (hpi (n + 1)) - PI)
     < (21 * INR (n + 1) + 3) /IZR magnifier.
\end{verbatim}
In the end, we obtain a description of the algorithm based on
integers, which can be applied to any number of iterations and any
suitable scaling factor.  This algorithm can already be used to
compute approximations of \(\pi\) inside Coq, but it will not return
answers in reasonable time for precisions that go beyond a thousand
digits (less than a second for a 7 iterations at 100 digits, 12
seconds for 9 iterations at 500 digits, a minute for 10
iterations at 1000 digits).

Concerning the magnitude of the accumulated error, for one million
digits the number of iterations is 20, and the error is guaranteed
to be smaller than 423.

\paragraph{Changing the scaling factor.} Although we are culturally
attracted
by the fractional representation of \(\pi\) in decimal form, it is more
efficient to perform most of the costly computations using a scaling
factor that is a power of 2.  For any two scaling factors \(m_1\) and
\(m_2\), let us assume that \(v_1\) and \(v_2\) are linked by the equation
\[v_2 =\left\lfloor \frac{v_1 \times m_2}{m_1}\right\rfloor.\]
If \(v_1\) is the representation of a constant \(a\) for the scaling
factor \(m_1\), then \(v_2\) is a reasonably good approximation of
\(a\) for the scaling factor \(m_2\).  This suggests that we could
perform all operations with a scaling factor \(m_1\) that is a power
of 2 and then post-process the result to obtain a representation for the
scaling factor \(m_2\).  Of course, one more multiplication and one
more division need to be performed and a little precision is lost in the
process, but the gain in computation time is worth it.

The validity of this change in scaling factor is expressed by the
following lemma.
\begin{verbatim}
Lemma change_magnifier : forall m1 m2 x, (0 < m2)%Z ->
  (m2 < m1)%Z ->
  hR m1 x - /IZR m2 < hR m2 (x * m2/m1) <= hR m1 x.
\end{verbatim}
This lemma expresses that the added error for this operation is
only one time the inverse of the new scaling factor.

In our case, we use this lemma with \({\tt m1} = 2 ^ {3321942}\) and
\({\tt m2} = 10^{10 ^6+4}\) for instance.
\paragraph{Guaranteeing a fixed number of digits.}
When we want to compute a number \(N\) of digits, we don't know in
advance whether the digits at position \(N+1\), \(N+2\), \dots describe
a small number or a large number.  If this number is too small or too
large we are unable to guarantee the value of the digit at position \(N\).

Let's illustrate this problem on a small example.  Let's assume we want
to compute the integral part of \(a\) and we have an approximated value
\(b\) which
is guaranteed to be within \(1/4\) of \(a\).  Moreover, when computing
\(b\) with a precision of 2 digits, we know that our computation process
may introduce errors of two {\em units in the last place}.  This means that
we actually compute a value \(c\) whose distance to \(b\) is guaranteed
to be smaller than \(0.02\).  At the time we discover the result of computing
\(c\) three cases may occur.
\begin{enumerate}
\item if the fractional part of \(c\) is smaller than \(0.27\), the
integral part of \(a\) may be smaller than the integral part of \(c\).
For instance, we may have \(c = 3.26\), \(b=3.245\), and \(a=2.995\)
\item if the fractional part of \(c\) is larger than or equal to
\(0.73\), the integral
part of \(a\) may be larger than the integral part of \(c\).  For
instance, we may have \(c=2.74\), \(b=2.755\), and \(a=3.005\).
\item if the fractional part of \(c\) is larger than or equal to \(0.27\)
or smaller than \(0.73\), then we now that \(a\), \(b\), and \(c\) all share
the same integral part.
\end{enumerate}
When considering distant decimals, the same problem is transposed through
multiplication by a large power of 10.

After putting
together the error coming from the difference \(\pi_n - \pi\), the
accumulated rounding errors, and the error coming from the change
of scaling factor, this means we need to verify that the last four
digits are either larger than 0427 or smaller than 9573.  This
verification is made in the following definitions, which return
a boolean value and a large integer.  The meaning of the two
values is expressed by the attached lemma.
\begin{verbatim}
Definition million_digit_pi : bool * Z :=
  let magnifier := (2 ^ 3321942)%Z in
  let n := hpi magnifier 20 in
    let n' := (n * 10 ^ (10 ^ 6 + 4) / 2 ^ 3321942)%Z in
    let (q, r) := Z.div_eucl n' (10 ^ 4) in
    ((427 <? r)%Z && (r <? 9573)%Z, q).

Lemma pi_osix :
  fst million_digit_pi = true ->
    hR (10 ^ (10 ^ 6)) (snd million_digit_pi) < PI <
    hR (10 ^ (10 ^ 6)) (snd million_digit_pi) +
    Rpower 10 (-(Rpower 10 6)).
\end{verbatim}
\paragraph{Proving the big number computations.}  The lemma
{\tt million\_digit\_pi} only states the correctness of computations
for computations in the type {\tt Z}, but this computation is
unpractical to perform.  The last step is to obtain
the same proof for computations on the type {\tt bigZ}.  The library
{\tt BigZ} provides both this type and a coercion function noted
{\tt [ \(\cdot\) ]} so that when {\tt x} is a big integer of type {\tt
  bigZ}, {\tt [x]} is the corresponding integer of type {\tt Z}.

In what follows, the functions {\tt rounding\_big.hmult}, et cetera
operate on numbers of type {\tt BigZ}, while the functions
{\tt hmult} operate on plain integers.  We have the following
morphism lemmas:
\begin{verbatim}
Lemma hmult_morph p x y:
  [rounding_big.hmult p x y] = hmult [p] [x] [y].
Proof.
unfold hmult, rounding_big.hmult.
rewrite BigZ.spec_div, BigZ.spec_mul; reflexivity.
Qed.

Lemma hdiv_morph p x y:
  [rounding_big.hdiv p x y] = hdiv [p] [x] [y].
Proof.
unfold hdiv, rounding_big.hdiv.
rewrite BigZ.spec_div, BigZ.spec_mul; reflexivity.
Qed.
\end{verbatim}
Using these lemmas, it is fairly routine to prove the
correspondence between the algorithms instantiated on both types.

\begin{verbatim}
Lemma hpi_rec_morph :
 forall s p n v1 v2 v3,
   [s] = hsqrt [p] (h2 [p]) ->
   [rounding_big.hpi_rec p n s v1 v2 v3] =
   hpi_rec [p] n [s] [v1] [v2] [v3].

Lemma hpi_morph : forall p n,
   [rounding_big.hpi p n]%bigZ = hpi [p]%bigZ n.
\end{verbatim}
In the end, we have a theorem that expresses the correctness of the
computations made with big numbers, with the following statement.

\begin{verbatim}
Lemma big_pi_osix :
  fst rounding_big.million_digit_pi = true ->
  (IZR [snd rounding_big.million_digit_pi] *
         Rpower 10 (-(Rpower 10 6)) <
    PI
   <
    IZR [snd rounding_big.million_digit_pi]
      * Rpower 10 (-(Rpower 10 6))
     + Rpower 10 (-(Rpower 10 6)))%R.
\end{verbatim}
This statement expresses that the computation returns a boolean value
and a large integer.  When this boolean value is {\tt true}, then
the large integer is the largest integer \(n\) so that
\[\frac{n}{10^{10^6}} < \pi.\]

The computation of this value takes approximately 2 hours on a
powerful machine.  We also implemented similar functions
to compute approximations of \(\pi\) using the Brent-Salamin
algorithm, and experiments showed the computation is twice as fast.
\section{Related work}
Computing approximations of \(\pi\) is a task that is necessary for many
projects of formally verified mathematics, but precision beyond tens
of digits are practically never required.  To our knowledge, this work is
the only one addressing explicitly the challenge of computing decimals at
position beyond one thousand.  Most developments rely on
Machin-like formulas to give a computationally relevant definition of \(\pi\).
The paper \cite{BertotAllais14} already provides an overview of methods
used to compute \(\pi\) in a variety of provers.  In Hol-Light
\cite{hol-light-analysis}, an
approximation to the precision of \(2^{-32}\) is obtained by approximating
\(\frac{\pi}{6}\) using the intermediate value theorem and a Taylor
expansion of the sine function, and the library also provides a description
of a variety of Machin-like formulas.  In
Isabelle/HOL \cite{Nipkow-Paulson-Wenzel:2002},
one of the Machin-like formulas is provided directly in the
basic theory of transcendental functions.  Computation of arbitrary
mathematical formulas, in the spirit of what is done with the {\tt interval}
tactic, is described in work by H\"olzl~\cite{Hoelzl09}.

The HOL Light library contains a formalization of the BBP formula~\cite{hol-bbp}.
Our contribution is to link the formalization of the formula with the actual algorithm
that computes the digit.

In the Coq system, real numbers can also be approached constructively as in
the C-CoRN library \cite{Cruz-Filipe04}.  This was used as the basis for a
library
providing fairly efficient computation of mathematical functions within
the theorem prover \cite{OConnor-2008,KrebbersSpitters}.  Using an advanced
Machin-like
formula they are capable to compute numbers like \(\sqrt{\pi}\) at a
precision of 500 digits in about 6 seconds (to be compared with less
than a second in our case, but our development is not as versatile as
theirs).

The formalized proof of the Kepler conjecture, under the supervision
of T. Hales \cite{Hales15} also required computing many inequalities
between mathematical formulas involving transcendental functions, a task
covered more specifically by Solovyev and Hales~\cite{SolovyevHalesNFM13}, but
none of these computations involved precisions in the ranges that we have been
studying here.
\section{Conclusion}
What we guarantee with our lemmas is that the integer we produce satisfies
a property with respect to \(\pi\) and a large power of the base, which
is 16 in the case of the the BBP algorithm, and may
be any integer in the case of the algebraic-geometric mean algorithms.  We do
not guarantee that the string produced by the Coq system when printing this
large number is correct, but experimental evidence shows that that part of the
Coq system (printing large numbers) is correct.  That computations can proceed
to the end is a nice surprise, because
it would be understandable that some parts of the theorem prover have
limitations that preclude heavy computing (as is the case when performing
computations with natural numbers, which are notoriously naive in their
implementation and their space and time complexity).  It would be an interesting
project to construct a formally verified integer to string converter, but
this project is probably not as challenging as what has been presented in this
article.

The organisation of proofs follows principles that were advocated by
Cohen, D\'en\`es, M\"ortberg, and Siles
\cite{refinement-algebra,refinements-free}, where the algorithm is first
studied in a mathematical setting using mathematical objects (in this
case real numbers) before being embodied in a more efficient implementation
using different data-types.  The concrete implementation is then viewed
as a refinement of the first algorithm.  This approach makes sure that
we take advantage of the most comfortable mathematical libraries when
performing the most difficult proofs.  The refinement approach was
used twice: first to establish the correspondence between computations
on real numbers and the computations on integers, and second to establish
the correspondence between integers and big integers.  The first stage
does not fit exactly the framework advocated by Cohen and co-authors,
because the computations are only approximated and we need to quantify
the quality of the approximation.  On the other hand, the second stage
corresponds quite precisely to what they advocate, and it was a source
of great simplification in our formal proof, because the Coq libraries
provided too few theorems and tactics to work on the big integers.

This experiment also raises the question of {\em what do we perceive
  as a formally verified program?}  The implementations described in
this paper do run and produce output, however they need the whole
context of the interactive theorem prover.  We experimented with using
the extraction facility of the Coq system to produce stand-alone
programs that can be compiled with OCaml and run independently.  This
works, but the resulting program is one order magnitude slower than
what runs in the interactive theorem prover.  The reason is that the
{\tt BigZ} library exploits an ability to compute directly with
machine integers (numbers modulo \(2^{31}\)) 
\cite{ArmandGregoireSpiwackThery2010}, while the extracted
program still views these numbers as records with 31 fields, with no
shortcuts to exploit bit-level algorithmics.
This raises several
questions of trusted base: firstly, the Coq system with the ability to
exploit machine integers directly for number computations has a wider
trusted base (because the code linking integer computation with
machine integer computation needs to be trusted).  This first question
is handled in another published article by Armand, Gr\'egoire, Spiwack and
Th\'ery \cite{ArmandGregoireSpiwackThery2010}.  Secondly we also have
to trust the implementation of the {\tt native\_compute} facility,
which generates an {\tt OCaml} program, calls the OCaml compiler, and
then runs and exploits the results of the compiled program.  This
question is handled in another article by Boespflug, D\'en\`es and Gr\'egoire
\cite{full-throttle}.  Thirdly, we could also extract the algorithms
as modules to be interfaced with arbitrary libraries for large number
computations.  We would thus obtain implementations that would be partially
verified and whose guarantees would depend on the correct implementation
of the large number operations.  This is probably the most sensible approach
to using formally verified algorithms in the real world.

In the direction of formally verified programs, the next stage will be
to study how the algorithms studied in this article can be implemented
using imperative programming languages, avoiding stack operations and
implementing clever memory operations, such as re-using explicitly the
space of data that has become useless, instead of relying on a
general purpose garbage-collector.  Obviously, we would need to interface
with a library for large number computations in such a setting.  Such
libraries already exist, but none of them have been formally verified.
We believe that the community of formal verification will produce such
a formally verified library for large number computations, probably
exploiting the advances provided by the CompCert formally verified compiler
\cite{Leroy-Compcert-CACM} (which provides the precise language for the
implementation),
and the Why3 tool \cite{filliatre13esop} to organize proofs of programs with
imperative features,
based on various forms of Hoare logic.

In their current implementation, our algorithms run at speeds that are several
orders of magnitude lower than the same algorithms implemented by clever
programmers in heavy duty libraries like {\tt mpfr} \cite{mpfr}.  For now,
the algorithms for elementary operations are based on Karatsuba-like
divide-and-conquer approaches, with binary tree implementations of large
numbers, but it could be interesting to implement fast-Fourier-transform
based multiplication as suggested by Sch\"onhage and Strassen
\cite{SchoenhageStrassen71}
and observe whether this brings a significative
improvement in the computation of billions of decimals.

In spite of the fun with mathematical curiosities around the \(\pi\)
number, the real lesson of this paper is more about the current progress
in interactive theorem provers.  How much mathematics can be described
formally now?
How much detail can we give about computations?  How reproducible is this
experiment?

For the question on how much mathematics, it is quite satisfying
that real analysis becomes feasible, with concepts such as improper integrals,
power series, interchange between limits, with automatic tools to check
that mathematical expressions stay within bounds, but also with rigidities
coming from the limits of the automatic tool.  One of the rigidity that
we experienced is the lack of a proper integration of square roots in the
automatic tool that deals with equalities in a field.  This tool, named
{\tt field}, deals very well with equalities between expressions that contain
mostly products, divisions, additions and subtractions, but it won't simplify
expressions such as
\(\sqrt{\frac{1}{\sqrt{2}} + \frac{1}{\sqrt{2}}} - \sqrt{\sqrt{2}}\).
From a human user's
perspective, this rigidity is often hard to accept, because once the properties
of the square root function are understood, we integrate them directly in
our mental calculation process.

For the question on how much detail we can give about computations, these
experiments show that we can go quite far in the direction of reasoning
about computation errors.  This is not a novelty, and many other experiments
by other authors have been studying how to reason about floating point
computations \cite{BolMel11}.  This experiment is slightly different in
that it relies more on fixed point computations.

For the question on how reproducible is this experiment, we believe that
one should distinguish between the task of running the formalized proof and
the task of developing it.  For the first task, re-running the formal proof,
we provide a link to the sources of our developments \cite{PlouffeAGMSources},
which can be run
with Coq version 8.5 and 8.6 and precise versions of the libraries Coquelicot
and
Interval.  For the task of developing the
formal proof, this becomes a question at the edge of our scientific expertise,
but still a question that is worth asking.  In the long run, formally
verified mathematics should become practical to a wider audience thanks
to the availability of comprehensive and well-documented libraries such
as Coquelicot \cite{BLM15} or mathematical components
\cite{MathematicalComponents}.  However,
there are some aspects of the work that make reproducibility by less
expert users difficult.  For instance, it is often difficult to understand
the true limits of automatic tools and this form of rigidity may cause
users to lose a lot of time, for instance by mistaking a failure to
prove a statement with the fact that the statement could be wrong.  Another
example is illustrated with the use of filters in the Coquelicot library,
which requires much more advanced mathematical expertise than what would
be expected for an intermediate level library about real analysis.

\bibliographystyle{plain}
\bibliography{main}

\begin{thebibliography}{10}

\bibitem{AlmkvistBerndt88}
Gert Almkvist and Bruce Berndt.
\newblock {\em Gauss, Landen, Ramanujan, the arithmetic-geometric mean,
  ellipses, $\pi$, and the Ladies Diary (1988)}, pages 125--150.
\newblock Springer International Publishing, 2016.

\bibitem{CapesAGM95}
Anonymous.
\newblock Première composition de mathématiques'', concours externe de
  recrutement de professeurs certifiés, section mathématiques, 1995.

\bibitem{ArmandGregoireSpiwackThery2010}
Micha{\"{e}}l Armand, Benjamin Gr{\'{e}}goire, Arnaud Spiwack, and Laurent
  Th{\'{e}}ry.
\newblock Extending {C}oq with {I}mperative {F}eatures and {I}ts {A}pplication
  to {SAT} {V}erification.
\newblock In {\em Interactive Theorem Proving, First International Conference,
  {ITP} 2010}, volume 6172 of {\em LNCS}, pages 83--98. Springer, 2010.

\bibitem{BBP97}
David Bailey, Peter Borwein, and Simon Plouffe.
\newblock On the rapid computation of various polylogarithmic constants.
\newblock {\em Mathematics of Computation}, 66(218):903--913, 1997.

\bibitem{Bertot:CPP15}
Yves Bertot.
\newblock Fixed precision patterns for the formal verification of mathematical
  constant approximations.
\newblock In {\em Proceedings of the 2015 Conference on Certified Programs and
  Proofs}, pages 147--155. ACM, 2015.

\bibitem{BertotAllais14}
Yves Bertot and Guillaume Allais.
\newblock {Views of Pi: definition and computation}.
\newblock {\em {Journal of Formalized Reasoning}}, 7(1):105--129, October 2014.

\bibitem{RacineCarreeGMP}
Yves Bertot, Nicolas Magaud, and Paul Zimmermann.
\newblock A proof of {GMP} square root.
\newblock {\em Journal of Automated Reasoning}, 29(3-4):225--252, 2002.

\bibitem{PlouffeAGMSources}
Yves Bertot, Laurence Rideau, and Laurent Th\'ery.
\newblock Distant decimals of pi, 2017.
\newblock Available at
  \url{https://www-sop.inria.fr/marelle/distant-decimals-pi/}.

\bibitem{DBLP:conf/types/Besson06}
Fr{\'e}d{\'e}ric Besson.
\newblock Fast reflexive arithmetic tactics the linear case and beyond.
\newblock In {\em Proceedings of the 2006 International Conference on Types for
  Proofs and Programs}, volume 4502 of {\em LNCS}, pages 48--62. Springer,
  2007.

\bibitem{full-throttle}
Mathieu Boespflug, Maxime D{\'e}n{\`e}s, and Benjamin Gr{\'e}goire.
\newblock Full reduction at full throttle.
\newblock In {\em Certified Programs and Proofs: First International
  Conference}, volume 7086 of {\em LNCS}, pages 362--377. Springer, 2011.

\bibitem{BLM15}
Sylvie Boldo, Catherine Lelay, and Guillaume Melquiond.
\newblock Coquelicot: A user-friendly library of real analysis for {C}oq.
\newblock {\em Mathematics in Computer Science}, 9(1):41--62, March 2015.

\bibitem{BolMel11}
Sylvie Boldo and Guillaume Melquiond.
\newblock Flocq: A unified library for proving floating-point algorithms in
  {C}oq.
\newblock In {\em IEEE Symposium on Computer Arithmetic}, pages 243--252. IEEE
  Computer Society, 2011.

\bibitem{BorweinAGM}
Jonathan~M. Borwein and Peter~B. Borwein.
\newblock {\em Pi and the AGM: A Study in the Analytic Number Theory and
  Computational Complexity}.
\newblock Wiley-Interscience, New York, NY, USA, 1987.

\bibitem{Brent76}
Richard~P. Brent.
\newblock Fast multiple-precision evaluation of elementary functions.
\newblock {\em J. ACM}, 23(2):242--251, April 1976.

\bibitem{Filtres-cartan}
Henri Cartan.
\newblock Th\'eorie des filtres.
\newblock {\em Comptes Rendus de l'Acad\'emie des Sciences}, 205:595--598,
  1937.

\bibitem{refinements-free}
Cyril Cohen, Maxime D{\'e}n{\`e}s, and Anders M{\"o}rtberg.
\newblock Refinements for free!
\newblock In {\em Certified Programs and Proofs: Third International
  Conference}, volume 8307 of {\em LNCS}, pages 147--162. Springer, 2013.

\bibitem{Cruz-Filipe04}
Luis Cruz-Filipe.
\newblock {\em Constructive Real Analysis: a Type-Theoretical Formalization and
  Applications}.
\newblock PhD thesis, University of Nijmegen, April 2004.

\bibitem{refinement-algebra}
Maxime D{\'{e}}n{\`{e}}s, Anders M{\"{o}}rtberg, and Vincent Siles.
\newblock A refinement-based approach to computational algebra in {C}oq.
\newblock In {\em Interactive Theorem Proving - Third International Conference,
  {ITP} 2012}, volume 7406 of {\em LNCS}, pages 83--98. Springer, 2012.

\bibitem{coq}
Coq development team.
\newblock The {C}oq proof assistant, 2016.
\newblock \url{http://coq.inria.fr}.

\bibitem{filliatre13esop}
Jean-Christophe Filli{\^a}tre and Andrei Paskevich.
\newblock Why3 --- where programs meet provers.
\newblock In {\em Programming Languages and Systems: 22nd European Symposium on
  Programming, ESOP 2013}, volume 7792 of {\em LNCS}, pages 125--128. Springer,
  2013.

\bibitem{mpfr}
Laurent Fousse, Guillaume Hanrot, Vincent Lef\`{e}vre, Patrick P{\'e}lissier,
  and Paul Zimmermann.
\newblock Mpfr: A multiple-precision binary floating-point library with correct
  rounding.
\newblock {\em ACM Trans. Math. Softw.}, 33(2), June 2007.

\bibitem{MathematicalComponents}
Georges Gonthier et~al.
\newblock Mathematical components.
\newblock Available at \url{http://math-comp.github.io/math-comp/}.

\bibitem{Gourevitch99}
Boris Gourevitch.
\newblock The world of {P}i, 1999.
\newblock Available at \url{http://www.pi314.net}, consulted March 2017.

\bibitem{GregoireTheryIJCAR2006}
Benjamin Gr{\'e}goire and Laurent Th{\'e}ry.
\newblock A purely functional library for modular arithmetic and its
  application to certifying large prime numbers.
\newblock In {\em Automated Reasoning: Third International Joint Conference,
  IJCAR 2006}, volume 4130 of {\em LNCS}, pages 423--437. Springer, 2006.

\bibitem{Hales15}
Thomas~C. Hales et~al.
\newblock A formal proof of the {K}epler conjecture.
\newblock {\em arXiv}, 1501.02155, 2015.

\bibitem{hol-light-analysis}
John Harrison.
\newblock Formalizing basic complex analysis.
\newblock In {\em From Insight to Proof: Festschrift in Honour of Andrzej
  Trybulec}, volume 10(23) of {\em Studies in Logic, Grammar and Rhetoric},
  pages 151--165. University of Bia{\l}ystok, 2007.
\newblock \url{http://mizar.org/trybulec65/}.

\bibitem{hol-bbp}
John Harrison.
\newblock Pi series in {B}ailey/{B}orwein/{P}louffe \textit{polylogarithmic
  constants} paper, the {HOL} {L}ight library, 2010.
\newblock Available at
  \url{https://github.com/jrh13/hol-light/blob/master/Examples/polylog.ml}.

\bibitem{Harrison98}
John Harrison.
\newblock {\em Theorem Proving with the Real Numbers}.
\newblock Springer Publishing Company, Incorporated, 1st edition, 2011.

\bibitem{Hoelzl09}
Johannes H{\"o}lzl.
\newblock Proving inequalities over reals with computation in {Isabelle/HOL}.
\newblock In {\em Proceedings of the {ACM SIGSAM 2009} International Workshop
  on Programming Languages for Mechanized Mathematics Systems ({PLMMS'09})},
  pages 38--45, Munich, 2009.

\bibitem{Isabelle-filters}
Johannes H{\"o}lzl, Fabian Immler, and Brian Huffman.
\newblock Type classes and filters for mathematical analysis in {Isabelle/HOL}.
\newblock In {\em Interactive Theorem Proving (ITP 2013)}, volume 7998 of {\em
  LNCS}, pages 279--294. Springer, 2013.

\bibitem{King1924}
Louis~V. King.
\newblock {\em On the Direct Numerical Calculation of Elliptic Functions and
  Integrals}.
\newblock Cambridge University Press, 1924.

\bibitem{KrebbersSpitters}
Robbert Krebbers and Bas Spitters.
\newblock Type classes for efficient exact real arithmetic in {C}oq.
\newblock {\em Logical Methods in Computer Science}, 9(1), 2011.

\bibitem{Leroy-Compcert-CACM}
Xavier Leroy.
\newblock Formal verification of a realistic compiler.
\newblock {\em Commun. ACM}, 52(7):107--115, July 2009.

\bibitem{interval}
{\'{E}}rik Martin{-}Dorel and Guillaume Melquiond.
\newblock Proving tight bounds on univariate expressions with elementary
  functions in {C}oq.
\newblock {\em Journal of Automated Reasoning}, 57(3):187--217, 2016.

\bibitem{Nipkow-Paulson-Wenzel:2002}
Tobias Nipkow, Markus Wenzel, and Lawrence~C. Paulson.
\newblock {\em Isabelle/HOL: A Proof Assistant for Higher-order Logic}.
\newblock Springer, Berlin, Heidelberg, 2002.

\bibitem{OConnor-2008}
Russell O'Connor.
\newblock Certified exact transcendental real number computation in {C}oq.
\newblock In {\em Theorem Proving in Higher Order Logics: 21st International
  Conference, TPHOLs 2008}, volume 5170 of {\em LNCS}, pages 246--261.
  Springer, 2008.

\bibitem{Salamin76}
Eugene Salamin.
\newblock Computation of \(\pi\) using arithmetic-geometric mean.
\newblock {\em Mathematics of Computation}, 33(135), 1976.

\bibitem{SchoenhageStrassen71}
Arnold Sch{\"{o}}nhage and Volker Strassen.
\newblock Schnelle {M}ultiplikation gro{\ss}er {Z}ahlen.
\newblock {\em Computing}, 7(3-4):281--292, 1971.

\bibitem{SolovyevHalesNFM13}
Alexey Solovyev and Thomas~C. Hales.
\newblock Formal verification of nonlinear inequalities with {T}aylor interval
  approximations.
\newblock In {\em NASA Formal Methods: 5th International Symposium, NFM 2013,},
  volume 7871 of {\em LNCS}, pages 383--397. Springer, 2013.

\end{thebibliography}

\end{document}